\documentstyle[preprint,epsf,aps]{revtex}

\newcommand{\bi}{\bibitem}
\newcommand{\be}{\begin{eqnarray}}
\newcommand{\ee}{\end{eqnarray}}
\newcommand{\nn}{\nonumber}
\catcode`\@=11
\def\lsim{\mathrel{\mathpalette\@versim<}}
\def\gsim{\mathrel{\mathpalette\@versim>}}
\def\@versim#1#2{\vcenter{\offinterlineskip
\ialign{$\m@th#1\hfil##\hfil$\crcr#2\crcr\sim\crcr } }}
\catcode`\@=12

\begin{document}

\draft
\preprint{KANAZAWA-00-05, UTCCP-P-87}
\title{A Study on \\the Non-perturbative Existence of Yang-Mills Theories\\
with\\
Large Extra Dimensions
}
\author{ Shinji Ejiri}
\address{
Center for Computational Physics, University
of Tsukuba, Tsukuba 305-8577, Japan
}
\author{Jisuke Kubo and Michika Murata}
\address{
Institute for Theoretical Physics, 
Kanazawa  University, 
Kanazawa 920-1192, Japan
}
\maketitle
\begin{abstract}
Pure lattice $SU(2)$ Yang-Mills
theory in five dimensions  is considered,
where an extra
dimension is  compactified on
a circle.  Monte-Carlo  simulations indicate that
the theory possesses  a continuum limit 
with a non-vanishing string tension if the 
compactification radius is smaller than 
 a certain value $R_M$ which is $O(1/10)$ of
the inverse of the square root of  the string tension.
We verify non-perturbatively the power-law running of gauge
coupling constant.
Our method can be applied to the investigation of 
continuum limits in  other 
higher-dimensional  gauge theories.

\end{abstract}
\pacs{11.10.Hi, 11.10.Kk,  11.10.Wx,  11.15.Ha, 11.25.Mj, 12.38.Gc}

\narrowtext

\section{INTRODUCTION}

The idea of unifying fundamental forces by introducing
extra dimensions has attracted attention for
many decades, and a theory realizing this 
idea is called  Kaluza-Klein theory  \cite{kaluza}.

Recently, it has been observed  by Arkani-Hamed, 
Dimopoulos and Dvali \cite{arkani1} that the existence
of extra dimensions may play an important r\^ ole
to understand the hierarchical scales that exist between 
the weak and  Planck scales. From a simple setting that only the graviton
can propagate in the bulk 
corresponding to  the extra dimensions
while all the other fields
of the Standard Model (SM) are located on
a four-dimensional wall,
they have concluded \cite{arkani1} that the length scale of 
the extra dimensions 
can be rather large $\gsim 10^{-2}$ cm, in contrast to 
  previously suggested
Kaluza-Klein theories in which the size of 
extra dimensions was  of the order
of the (four-dimensional)  Planck length $10^{-33}$ cm or 
 $1/M_{\rm GUT} \approx 10^{-30}$ cm, 
where $M_{\rm GUT}$ is the unification scale
in four-dimensional grand unified theories (GUTs).
Their idea has been then followed and extended by several
authors \cite{randall1,cohen1} to  obtain more satisfying solutions of 
the hierarchy problem.
Moreover, the above phenomenological
proposal to confine fields on a lower-dimensional 
subspace
fits well  \cite{witten1,antoniadis4,dienes1,kakushadze1} 
the D-branes
\cite{polchinski1} (extended objects attached by 
the end points of open strings) in  string theories.

If part of the SM fields can propagate in the bulk,
and the size of the extra dimensions are large,
the existence of such extra
dimensions may be experimentally verified.
There will be a number of phenomenological questions  
(see \cite{giudice1} for instance)
like
`` what are the experimental bounds on the size of
the extra dimensions? \cite{accomando}'' 
However,
our concern in this paper is of a 
theoretical nature: Is the existence of a large extra
dimension consistent with quantum theory?
Our answer to this question will be ``Yes'', provided that
the compactification radius $R$ is smaller than a
certain value, the maximal radius  $R_M$.
It should be emphasized that the previous investigations 
\cite{creutz1,okamoto1,nishimura1} on 
 non-abelian gauge theories in five dimensions
on a lattice 
(which indicated that the theory have no continuum limit)
were performed in the uncompactified case.
These works \cite{okamoto1,nishimura1} were
 motivated to investigate whether or not
the non-trivial ultraviolet
fixed point  found in the $\epsilon$-expansion
\cite{peskin} is real.

To be more specific, we consider pure $SU(2)$ Yang-Mills
theory in five dimensions where an extra
dimension is compactified on
a circle $S$ with the radius of $R$. 
(It would be more ``realistic''
 to compactly the 
fifth dimension on the orbifold $S/Z_2$ so that
the zero-modes contain only  four-dimensional gauge
fields and no scalar fields.
We leave the case of $S/Z_2$ to future work.)
One may expect that the theory will carry the
basic property of a four-dimensional gauge theory if the radius
$R$ is sufficiently small, while in the opposite limit
of $R$ the theory becomes more five dimensional.
So there may be the maximal radius $R_M$ below which
the theory can possess a continuum limit 
with a non-vanishing string tension and can exist non-perturbatively.
We will indeed  find that our numerical simulations based on
a compactified lattice gauge theory are supporting the correctness of
this heuristic picture.

The string tension is one of the most familiar physical quantities, 
which can give a physical scale to  the lattice spacing.
However, at a deconfining phase transition of first order, 
the string tension vanishes discontinuously, and 
we  cannot use it for that purpose in this case.
One of the crucial observations in this paper is that, 
if the fifth dimension is compactified,
the first order phase transition changes its nature at a certain 
compactification radius.
We will see this  on anisotropic lattices
by performing Monte-Carlo simulations with various 
compactification radii and by investigating the phase structure.
The simulations also indicate that it could be possible to give 
a physical scale to the lattice spacing
even in the deconfining phase if the theory is compactified,
and this possibility will be studied more in detail.

We will assume that the phase transition  due to the compactification occurs
at a certain value of $R$, the critical compactification radius $R_C$, 
and that the compactification radius is kept fixed at 
$R_C$ along the critical 
line of the phase transition  due to the compactification.
That is, the critical compactification radius $R_C$
is assumed to 
 be a physical quantity.
This assumption enables us to compute the lattice
$\beta$-function  for a given $R$ as a function of
the lattice spacing $a_4$ of the four-dimensional 
direction. In doing so, we can verify non-perturbatively 
the power-law running of the gauge
coupling constant $g$, and find that the observed power law
behavior fits well to the one-loop form suggested 
in Refs. 
\cite{dienes1,peskin,veneziano,ross1,clark,mohapatra,kakushadze2,ross2,kubo2}. 
The results for  the lattice
$\beta$-function  obtained from
our Monte-Carlo simulations 
indicate the self-consistency of the assumption above.

The results obtained for  the lattice
$\beta$-function  can also be used to 
make a further assumption on the physical scale 
in the deconfining phase and
to investigate various 
scaling properties of the longitudinal Creutz ratio
(defined in Eq. (\ref{creutz})), making 
a discussion on  the existence of  continuum limits of the theory 
 possible.
We will be led to the interpretation
that the theory may possess  a continuum limit 
with a non-vanishing longitudinal string tension if the 
compactification radius $R$ is smaller than $R_M \approx R_C/3$, and
that the non-trivial ultraviolet
fixed point  found in the $\epsilon$-expansion
in the continuum theory may  no longer be spurious. 

After we define our lattice action in Section II, we 
start to present the details of our calculations.
In Section III we calculate the ratio of the lattice spacings
$\xi=a_4/a_5$ in terms of the parameters of the simulations
$\beta$ and $\gamma$, and then  we discuss the phase structure
in Section IV. In Section V we compute the lattice
$\beta$-function and then study on a continuum limit in Section VI, 
and the last
Section is devoted for  conclusion.

\section{THE ACTION}

In order to investigate the effects of a compactification in the
5-dimensional  $SU(2)$ gauge theory, it is crucial to employ an anisotropic
lattice  which has different lattice spacings, $a_4$ and $a_5$,
in the four-dimensional directions and in the fifth direction,
and is often used 
in the case of lattice gauge theories at finite temperature. 
We find that the effects  of the compactification 
on an isotropic lattice
can appear only for a small lattice size of the fifth
direction $( \leq 2 )$ so that it is practically impossible to study 
the theory with different sizes of this direction.
Another advantage is that, 
since we can vary $a_4$ and $a_5$ independently,
we can investigate the $a_4$
dependence  of physical quantities while keeping $a_5$ fixed.
This enables us to study scaling properties 
in the compactifed theory 
for a given compactification radius $R$.

We denote the five-dimensional lattice coordinates by $z_M
~(M=1,\dots,5)$
while  the four-dimensional ones by $x_{\mu}~(\mu=1,\dots,4)$
and  the fifth one by $y$.
Link variable takes the form
\be
 U_{M}(x,y) &=& \{~ U_{\mu}(x,y)=U(x,y;x+a_4 \hat{\mu},y)~,~ 
U_{5}(x,y)=U(x,y; x,y+a_5) ~\}~,
\label{link1}
\ee
where $U(z_1;z_2) \in SU(N_C)$ is the parallel
transporter.
The plaquette variables are
\be
U_{P_4} &=& U_{\mu\nu}(x,y) = U_{\mu}(x,y)
 U_{\nu}(x+a_4 \hat{\mu},y) 
U_{\mu}^{\dag}(x+a_4 \hat{\nu},y) U_{\nu}^{\dag}(x,y)\nn\\
U_{P_5} &=& U_{\mu 5}(x,y) = U_{\mu}(x,y)
 U_{5}(x+a_4 \hat{\mu},y)
U_{\mu}^{\dag}(x,y+a_5) U_{5}^{\dag}(x,y)~.
\label{link2}
\ee
The Wilson action
for  pure $SU(N_C)$ Yang-Mills theory in five dimensions is given by
\be
S &=& \beta_4~\sum_{P_4}~[~
1-\frac{1}{N_C}~\mbox{Re Tr} ~U_{P_4}~]+
\beta_5~\sum_{P_5}~[~
1-\frac{1}{N_C}~\mbox{Re Tr}~ U_{P_5}~]~,
\label{action}
\ee
where $\sum_{P_4}=\sum_{z~1 \leq \mu < \nu \leq 4}$ and
$\sum_{P_5}=\sum_{z~1 \leq \mu \leq 4}$.
Periodic boundary conditions are imposed in all the directions.
\footnote{
Another interesting case, i.e., orbifold boundary conditions which kill the
scalar zero mode, can be archived by imposing
$U(x,y;x,y+a_5)=U^{\dag}(x,-y-a_5;x,-y)$.
}
The coupling- and
correlation-anisotropy parameters are defined as
\be
\gamma &=& \sqrt{\frac{\beta_5}{\beta_4}}~,~\xi = \frac{a_4}{a_5}~,
\label{gamma-xi}
\ee
where $\gamma =\xi$ is satisfied in the tree level. 
In the naive continue limit $a_4, a_5 \to 0$ with
the length of the fifth dimension fixed at $2\pi R$,
the action (\ref{action}) becomes
\be
S &=& -\sum_{x,y} ~
[~(\frac{\beta_4 a_4^4}{2 N_C} )~
\frac{1}{2}\mbox{Tr}F_{\mu\nu}^2+
~(\frac{\beta_5 a_4^2 a_5^2}{2 N_C} )~
\mbox{Tr}F_{5\nu}^2~]+O(a^5)~,
\label{action2}
\ee
which goes to
\be
\int d^4 x \int_{0}^{2\pi R}d y ~\frac{-1}{2g_5^2}~\mbox{Tr}F_{M N}^2~,
\ee
if $
\beta_4 = 2 N_C a_5/g_5^2 $ and $
\beta_5 =2 N_C a_4^2/g_5^2 a_5$,
where
$
A_M =g_5 A_M^a T^a~,~
F_{MN}=\partial_M A_N-\partial_N A_M
-i [~A_M~,~A_N~]$,
and we have used
\be
 U_{\mu}(x,y) &=&e^{ ig_5a_4 A_{\mu} (x,y) }~,~
U_{5}(x,y)=e^{i g_5 a_5 A_{5}(x,y)}~.
\ee

On a lattice we mean a compactification if
\be
\frac{N_4 a_4}{N_5 a_5} & = &
\frac{a_4 N_4}{2\pi R} =\frac{N_4}{N_5}~\xi > 1
\ee
is satisfied.
Note that the gauge coupling constant $g_5$ has the dimension
of $\sqrt{a_4}$, and can be expressed as
\be
g_5^{-2} &=& \frac{\beta}{2 N_C a_4}~,~\beta =\sqrt{\beta_4 \beta_5}~.
\label{g5}
\ee
Later on we will use a dimensionless coupling constant $g$,
\be
g^{-2} &=& (2\pi R) g_5^{-2}=\frac{N_5 \beta}{2 N_C \xi}~,
\label{g-def1}
\ee
which is normalized for the four-dimensional
Yang-Mills theory with the tower of the Kaluza-Klein excitations.
At this point, Eq. (\ref{g-def1}) is only a tree-level definition.

\section{$\xi$--$\gamma$ RELATION}

The parameters of the simulations are $\beta$ and $\gamma$ for a 
given size of lattice, and the lattice spacings $a_4$ and $a_5$ are 
functions of these parameters.
The introduction of an anisotropy into a lattice means that
the regularization breaks  $O(5)$ invariance of the 
continuum theory. To recover this symmetry we have to fine
tune the anisotropy  parameters $\gamma$ and $\xi$ that 
are defined in Eq. (\ref{gamma-xi}).
At the tree-level, it is $\xi=\gamma$ as we have 
stated in the previous
section. In higher orders the tree-level relation suffers from  quantum
corrections so that
it can depend on $\beta$ and $\gamma$, i.e.,
$\xi=\xi(\gamma,\beta)$. 
The basic idea to find the corrected 
relation, which has been intensively used in the study of QCD at finite
temperature,  is to use that symmetry.
 There are variants of the method, and we
have decided to use a slightly modified  method that is based on the
matching of the Wilson loop
ratio \cite{burgers,klassen,scheideler}. 
Let us briefly explain the method below.

We consider  two kinds of Wilson loops $W(z_M,z_N)$, the one
$W(x_\mu,x_\nu)$ within the 
four-dimensional subspace and the other one $W(x_\mu,y)$ that
is extended into the fifth dimension, and calculate the ratios
\be
R(x_\mu,x_\nu)
&=&\frac{W(x_\mu+a_4 \hat{\mu},x_\mu)}{W(x_\mu,x_\nu)}~~\mbox{and}~~
R(x_\mu,y)
=\frac{W(x_\mu+a_4 \hat{\mu},y)}{W(x_\mu,y)}\ .
\label{ratios}
\ee
Since the Wilson loop is related to the static quark potential as
\be
W(z_M,z_N) \sim \exp\{-z_M V(z_N)\}~~  
\mbox{ ~~~for $z_M \rightarrow \infty $}~,
\ee 
we find that the rations (\ref{ratios})
for large $x$ and $y$ become
\be
R(x_\mu,x_\nu) 
&\sim & \exp\{ -a_4 V(x_\nu)\}\ ,~R(x_\mu,y)
\sim \exp\{ -a_4 V(y)\} \ .
\ee
The $O(5)$ symmetry of the continuum theory requires then that 
\be
R(x_\mu,x_\nu)=k~R(x_\mu,y) ~~\mbox{for}~~
x_\nu=n_\nu a_4=y=n_5 a_5~,
\label{xi-princ}
\ee
where we have allowed the presence of the factor $k$. 
We measure the ratios for a given set of the lattice size,
$\beta$ and $\gamma$, and assume that they take the form
\be
R(x_\mu,a_4 n_\nu)
&\sim & k_1\exp\{-\sigma_4 n_4\} ~~\mbox{and}~~
R(x_\mu,a_5 n_5)
\sim k_2\exp\{-\sigma_5 n_5\} \ ,
\label{ratios1}
\ee      
and that they should become identical with each other, by symmetry, when
$n_\nu a_4=n_5 a_5$.  From this consideration we obtain 
$\xi = a_4/a_5=\sigma_4/\sigma_5$.
Note that the ansatz (\ref{ratios1})
has a meaning  only in the confining region
of the parameters, of course.

In the practice, we fit the ansatz (\ref{ratios1}) for the data,
and then
scale $n_5$ by $z$ (i.e. $n_5 \to z n_4$)
in such a way that $R(x_\mu,z a_5 n_4)$ becomes
closest to $R(x_\mu,a_4 n_4)$, where we assume that $k=1$
on the r.h.side of 
Eq. (\ref{xi-princ}) \footnote{On a lattice where one
can obtain more data points, it is more convenient to
use the method developed in Ref. \cite{scheideler}
for QCD,
in which $k$ is different from $1$. 
In our case, due to the size of our lattice, we cannot obtain
enough number of data points. In such case  $k=1$
is a  reasonable assumption, as it has been discussed in
Ref. \cite{klassen}.}.
In the ideal
case we would have $z=\sigma_1/\sigma_2=\xi$.

To restore the $O(5)$ symmetry
in an efficient way,
simulations are performed using the heat bath algorithm 
of the $SU(2)$ lattice gauge theory.
We employ a lattice of $ N_4^4 \times N_5$ shown in TABLE I, where  
$N_5 \sim \gamma N_4 $ is satisfied.
We generate 5000 configurations, and
Wilson loops are measured every 5 configurations.

Fig. 1 shows $\xi$ versus $\beta$ for various values
of $\gamma$, and  we see that $\xi$ is almost independent
of $\beta$. The data points for larger $\beta$ are not plotted
because they correspond to the deconfining region
so that the ansatz (\ref{ratios1}) has no meaning.
The same data are plotted in Fig. 2 which shows the $\gamma$ dependence
of $\xi$. 
The data are summarized in TABLE I.
The central value of $\xi$ in the table is the average of
the data points in Fig. 1 for a fixed $\gamma$.

\section{PHASE STRUCTURE}

In this section we would like to investigate the phase structure of 
the five-dimensional theory defined by the action (\ref{action}). 
It is known from the mean field analysis that  higher-dimensional 
lattice gauge theories in more than four dimensions
have a first order phase transition 
\footnote{See for instance Ref. \cite{itzykson} and
references therein.}.
The studies of Monte-Carlo simulations 
\cite{creutz1,okamoto1} also indicate that
 in the case of  $SU(2)$ gauge theory the first order transition occurs 
starting at $D=5$. 
Our task is to extend these analyses to the compactified theory.
To this end,  we will be intensively using anisotropic lattices to
take into account the compactification of the fifth dimension.

\subsection{Longitudinal Creutz ratio}

The string tension between two quarks 
that are separated in space
is a typical physical quantity for the theory. 
What we know from experiments is that 
the string tension $\sigma_{\rm phys}$  between two quarks 
that are separated in the four-dimensional subspace
should be non-vanishing so that the potential
between them is linearly increasing with the distance $r$.
And the string tension is a good physical quantity to introduce 
a physical scale for other quantities obtained by a lattice calculation. 
If the underlying  gauge theory is formulated in 
five dimensions, however,
the feature of the linearly increasing potential is not automatically
present, and in fact, the first order deconfining transition is 
found in Refs. \cite{creutz1,okamoto1}. 

We measure the Creutz ratio $\chi(i,j)$ defined as
\be
\chi (i,j) &=&-\ln \{\frac{W(i,j)W(i-1,j-1)}{W(i,j-1) W(i-1,j)}\}~ , 
\label{creutz}
\ee
where $W(i,j)$ is a rectangular Wilson loop with lengths of $i$ and $j$.
The Creutz ratio with large $i$ and $j$ 
becomes the lattice string tension $\sigma_{\rm lat}$
in the case of the linearly increasing potential between two quarks.
So, if a Creutz ratio with large $i$ and $j$ takes a non-zero value, 
the corresponding Wilson loop shows the area law 
which we regard  as ``confinement''.
We consider the Wilson loops longitudinal to 
the four-dimensional subspaces,
because we are interested in the
confinement property in this subspace.
We would like to demonstrate that the Creutz ratio
behaves differently for different types of lattice.
The results obtained from Monte-Carlo simulations 
on an isotropic lattice of size $8^5$  ($\gamma^2=1.0$) and 
on an anisotropic lattice of the same size
($\gamma^2 = 2.0$ and $\gamma^2 = 4.0$)
are shown in Fig. 3, 
where 
the vertical axis  stands 
for the Creutz ratio, and the horizontal axis stands for
$\beta =  \sqrt{\beta_4 \beta_5}$.
We have generated 2500 configurations for each simulation point 
after thermalization, and 
Wilson loops are measured every 5 configurations for the calculation 
of the Creutz ratio.

We see from Fig. 3
that the phase transition between the confining and deconfining phase 
exists around $\beta = 1.64$ in the case of the isotropic 
lattice ($\gamma^2=1.0$) as it was found in Refs. \cite{creutz1,okamoto1}
and around $\beta \approx 1.73$ and $1.77$ in the cases of 
$\gamma^2 =2.0$ and $4.0$, respectively.
We have performed the simulations starting with an ordered configuration with 
$U_{M} = 1$ (defined in Eq. (\ref{link1}))
and with  a disordered configuration,
thereby obtaining  clear hysteresis curves.
The open symbols are the results of the ordered start and 
the filled symbols are those of the disordered start.
Our results indicate that the transitions are of first order,
in accord with the finding of 
Refs. \cite{creutz1,okamoto1} for $\gamma^2=1.0$.

\subsection{Transverse Polyakov loop}
In the uncompactified case, the Polyakov loop plays 
the r\^ ole for an indicator of confinement.
Here we consider 
loops which  are transverse
to the four-dimensional subspace
and define the transverse Polyakov loop as
\be
L &= & z \frac{1}{N_4^4} \sum_{x} \frac{1}{N_{C}} 
{\rm Tr} \prod_{y} U_5(x,y )~,
\label{polyakov}
\ee
where $z$ is a $Z(N_C)$ phase factor ($z^{N_C} = 1$) 
such that $\arg(L) \in (-\pi/N_C,\pi/N_C)$. 
In contrast to the longitudinal Creutz ratio (\ref{creutz})
which we have discussed in the previous subsection, 
the transverse Polyakov loop (\ref{polyakov})
has no direct physical meaning in four dimensions,
because we do not identify the fifth
direction with the temporal direction.
We may say however that 
the quark currents running into the fifth direction are confined if
the transverse Polyakov loop $\langle L \rangle$ vanishes. 

Fig. 4 shows  the results of the transverse 
Polyakov loop on the $8^5$ and 
$8^4 \times 12$ lattices for various values of $\gamma$, while,
 for comparison, the average of the plaquettes ($1 \times 1$ Wilson loop) 
for the
same lattices is shown in Fig. 5.
2500 configurations have been used to measure the Polyakov loop 
and the plaquette for each point.
As in the previous subsection, 
the open symbols are the results of the ordered start and 
the filled symbols are those of the disordered start.
As expected, we obtain  clear hysteresis curves,
and so the transverse  Polyakov loop
and the  average of the plaquettes also
indicate that the phase transition is of first order.

\subsection{Compactification effects}

It may be worth pointing out that the compactified 
$(D+1)$-dimensional $SU(N_C)$ lattice gauge theory  belongs to
the same universality class as
the $D$-dimensional $Z(N_C)$ spin model.
The case of QCD 
at finite temperature $T$ is a well-known example, where
the temporal direction is
compactified with the length $T^{-1}$.
We expect the existence of a similar  phase transition due to the 
compactification in our case, which is of second order, because
the phase transition in the four-dimensional $Z(2)$ spin model (Ising model) 
is of second order. 
So, we repeat the measurements of the transverse Polyakov
loop  (\ref{polyakov}) and 
the average of plaquette for the compactified case.

In order to  take into account the compactification of the fifth-dimension,
we use anisotropic lattices of size $8^4\times 4$ and $8^4\times 6$.
The results for the transverse Polyakov loop 
with different $\gamma$ are shown in Fig. 6 and 7. 
(In Fig. 6 we have included the result on a $12^4 \times 4$ lattice
which shows  that there are practically no finite size effects.)
Noticing  that the compactification radius $R$($=N_5 a_5/2\pi$)
becomes  smaller for a given $N_5$ as $\gamma$ becomes larger 
(see  Fig. 2 and TABLE I), we observe that the nature of the
phase  transition  changes due to the compactification.
Namely, the interval of $\beta$ in which two phases coexist 
becomes narrower as  $\gamma$ increases, 
and there are no intervals
for $\gamma^2 \gsim 2$ for the $8^4 \times 4$ case and 
for $\gamma^2 \gsim 4$ for the $8^4 \times 6$ case, respectively.
These phase transitions  seem to be of second order.
Observe also that the transition interval of $\beta$
for $\gamma^2=1.0$ does not depend on $N_5$, while,
in contrast to this, 
the transition point $\beta_C$ 
for the second order transition 
for a given $\gamma$ depends on $N_5$.  From these results, we  conclude that 
the second order phase transition 
is caused by the compactification, and that
the first order transition is not related to the compactification.
In Figs. 8 and 9, we plot  the average of the plaquettes 
for the $8^4\times 4$ and $8^4\times 6$ lattices. 
The results show that the transition becomes weak 
(like a cross over transition)
starting  at $\gamma$ at which 
the first order transition of the transverse 
Polyakov loop turns to be of second
order.  (In Fig. 8 we have included the result on a $12^4 \times 4$ lattice
to make it sure that finite size effects are negligible.)

In Fig. 10 we show the qualitative nature of the phase structure
in the $\beta_4-\beta_5$ plane, which we have obtained from the result of this
section.
The ``confining'' and ``deconfining'' phases are separated
by the critical lines of the first and second order phase transitions.
The position of the critical line (bold line) of the 
first order phase transition does not depend on the lattice size,
while that of the second order one  (solid line) depends crucially on 
$N_5$. Below the critical line in the  $\beta_4-\beta_5$ plane,
the transverse Polyakov loop vanishes, and it is different from zero above the
line. Note that this does not necessarily mean that the longitudinal
Creutz ratio (\ref{creutz}) vanishes in the
deconfining phase. 
The longitudinal
Creutz ratio (\ref{creutz})  corresponds to 
the "spatial string tension" in  QCD at finite temperature,
which is  defined by the spatial Wilson loop, 
and indeed
is non-vanishing even in the deconfining phase \cite{bali}.
Fig. 11 shows 
the  longitudinal Creutz ratio versus $\beta$ for the anisotropic lattice
of size $8^4\times 4$ with $\gamma^2$ fixed at $4.0$.
The figure  shows that the longitudinal Creutz
ratio varies smoothly as $\beta$ 
enters into the deconfining phase of the transverse Polyakov
loop,  indicating that
it could be possible to give 
a physical scale to the lattice spacing
even in that  phase.
Since indeed the spatial string tension is known to obey a scaling law
 at  high temperature \cite{bali},
we may wonder whether
some continuum limit in the present might also exist.
The following subsections and sections are
devoted to investigate
this possibility from another point of view.

In the case of QCD 
at finite temperature, the critical  temperature  $T_C$
is a physical quantity.  As in that case,
it is well possible  that the critical compactification radius
$R_C$ is a physical quantity, and that
the lattice system on the different critical lines 
in the $\beta-\gamma$ plane for different $N_5$ corresponds to
the same physical system.
As a first check, we estimate roughly the critical radius $R_C$
for two critical lines of the second order
phase transition at the end point.
As mentioned (see also Fig. 16), at $\gamma \approx  
\sqrt{2.0}$ for $N_5 = 4$ and 
at $\gamma \approx 2.0$ for $N_5 = 6$, 
the second order transition line  
merges in  the first order transition line. 
The value of  $\xi$ 
at the merging points, respectively,  is $1.78$ for $\gamma = \sqrt{2.0}$ 
and $2.78$ for $\gamma = 2.0$, where we have used the data in 
TABLE I. From the data on the Creutz ratio for the  $8^5$ lattice
(Fig. 3),  we find that 
the value of the longitudinal Creutz ratio at the transition points
is approximately constant
independent of $\gamma$, i.e.
\be
\sigma_{\rm lat} &=& \sigma_{\rm phys} a^2_4
\approx 0.7~, 
\label{sigma-lat}
\ee
where we identify the longitudinal Creutz ratio 
$ \chi(i,j)$ with large $i$ and $j$ as the 
lattice string tension $\sigma_{\rm lat} $.
Using this, we find
\be
R_C &=& \frac{N_5 a_{5c}}{ 2 \pi} \approx
\frac{N_5}{ 2 \pi\xi_c}~ \left[\frac{0.7}{\sigma_{\rm phys}}\right]^{1/2}
\approx
\left\{
\begin{array}{l}
0.30 / \sqrt{\sigma_{\rm phys}} \\
0.29 / \sqrt{\sigma_{\rm phys}}  \end{array}\right.
~~\mbox{  for}~~ \left\{
\begin{array}{l}
N_5=4 \\ N_5=6 \end{array}\right.~,
\label{rc}
\ee
where $\xi_c=a_4/a_{5c}$.
These values are consistent with 
the assumption that
the lattice system on the different critical lines corresponds to
the same physical system.
Eq. (\ref{sigma-lat}) also means that the value of 
$a_4$ at which the first order phase transition appears
is approximately independent of $\gamma$,
indicating that this value might have a
sensible meaning.
In the next section, we 
will do another check by using the lattice $\beta$-function.

\section{THE LATTICE $\beta$-FUNCTION}

We are interested in the physics in the four-dimensional subspace 
with a certain compactification radius.
The anisotropic lattice we have used in the previous section is 
convenient for computations with different $a_4$ 
while  keeping the compactification radius constant.
In this section we would like to compute the lattice
$\beta$-function in the four-dimensional subspace 
with the compactification radius $R$ fixed
at a certain value:
\be
\beta_{\rm lat} &=&
-a_4 \frac{d g^2}{d a_4} ~,
\label{beta1}
\ee
where $g=g_5/\sqrt{2\pi R}$ is the four-dimensional,
dimensionless gauge coupling.
We will calculate in the subsection B the $\beta$-function
for  the compactification radius $R$ at  the
critical compactification radius $R_C$
using two lattices with different $N_5$, where
$N_5$ also corresponds to the number of Kaluza-Klein 
excitations. So, if the theory we investigate should be regarded as 
a four-dimensional theory with only a few number of Kaluza-Klein excitations,
the $\beta$-function should depend explicitly on $N_5$. 
On the other hand, if we obtain the same lattice $\beta$-function for 
different $N_5$, 
we are indeed  dealing with a  five-dimensional theory,
and  finite $N_5$ or equivalently finite $a_5$ effects may
be regarded as negligibly small.
First we would like to check this point.
Another motivation is that we would like to examine
non-perturbatively  the celebrated power behavior of
the running of the gauge couplings in higher dimensions,
which we will use  to give a physical
scale in the deconfining phase of the
transverse Polyakov loop and then to
discuss the scaling behavior
of the longitudinal Creutz ratio (\ref{creutz})
in the next section.

Since the gauge coupling $g$ and the lattice $\beta$-function 
$\beta_{\rm lat}$ are dimensionless, we may assume that
the lattice spacings $a_4$ and $a_5$ enter only in the combination
$\xi=a_4/a_5$.
Furthermore, the perturbative analyses
and also the discussion that follows below suggest that
the correct variable is
\be
 s &\equiv & \frac{2\pi N_5}{ \xi} =\frac{2\pi N_5 a_5}{ a_4}
=\frac{(2\pi)^2 R}{ a_4}~.
\label{es}
\ee
This choice of the parameter has a non-trivial meaning:
We may conclude that, if $g$ really 
depends only on $s$,
the continuum limit $a_5 \to 0$ with 
the compactification radius $R$ fixed
can be taken, and 
 $R$ can be regarded as a physical quantity in this sense.

In the case of QCD at finite temperature,
the critical temperature $T_C$ is a universal quantity.
The analogy for our case would be that the
critical radius $R_C$ is a universal quantity of the
theory. So, the compactification radius would remain constant
along the critical line in the $\beta - \gamma$ plane.
However, there is a crucial difference compared with the case
of QCD  at finite temperature, because the critical lines
in the present case merge into the region of the
first order phase transition which is not related to the compactification.
Therefore, this assumption is not reliable in the region 
in which the transition is of the first order.

Keeping these circumstances in mind
and  defining  the gauge coupling as
\be
g^{-2} &=& 
2\pi R_C g^{-2}_{5} =\frac{N_5}{4}\frac{\beta_C}{\xi}~
\label{g-def}
\ee
on the critical line of the second order phase transition
\footnote{This definition of the gauge coupling
has the same form as the tree-level one (\ref{g-def1}).},
we can re-write 
 Eq. (\ref{beta1}) as
\be
\beta_{\rm lat} &=& \bar{\beta}_{\rm lat} ~(~
  1-\frac{d \ln R_C}{d \ln a_4}~)\nn\\
\mbox{with} & &
\bar{\beta}_{\rm lat}=
\frac{4}{N_5}~s \frac{d }{d s}
 \left[ ~ \frac{\xi }{\beta_C}(s) ~\right] =
- \frac{ 4\xi}{N_5\beta_C}~
 \left[ 1- \frac{\xi}{\beta_C}\frac{d\beta_C}{d\gamma }
 \frac{d\gamma}{d\xi }~\right] ~,
\label{beta2}
\ee
where use have been made of Eqs. (\ref{gamma-xi}), (\ref{g5})
and (\ref{g-def1}).
Here, we denote $\bar{\beta}_{\rm lat}$ for the $\beta$ function 
with the assumption that the $R_C$ is constant along the transition line.
If there is no $a_4$ dependence of $R_C$,
this assumption is correct so that 
$ \beta_{\rm lat}= \bar{\beta}_{\rm lat}$.

Note that the critical lines in the $\beta-\gamma$ plane 
are different for different $N_5$.
In Eq. (\ref{g-def}) we  are implicitly assuming that 
$g$ dose not depend on which critical line we use to calculate it.
If we obtain the same gauge coupling from the different lines,
it is a sign that the critical lattice systems for 
different $N_5$ describe the same physical system.
This will be checked in  subsection B.

\subsection{Precise determination of the critical lines}

To compute the lattice $\beta$ function $\bar{\beta}_{\rm lat}$
using Eq. (\ref{beta2}), we need to know precisely the location of the
critical points and its derivative with respect to $\gamma$ 
in the $\beta - \gamma$ plane.
Let us therefore determine the critical lines
in the $\beta-\gamma$ space next. 
To this end,  we identify
the transition point with the position of the peak 
of the susceptibility 
\be
\chi_L &=& N_4^4(\langle L^2 \rangle - \langle L \rangle^2)~,
\ee
where $L$ is the transverse Polyakov loop defined in (\ref{polyakov}).
We apply the histogram method \cite{swendsen} extended 
to an anisotropic lattice to evaluate the 
continuous parameter dependence 
of $\chi_L$,
as it was done in Ref. \cite{ejiri}.
To measure the Polyakov loop susceptibility,
we take 100,000 configurations. 
The results are plotted in Fig. 12 for $N_5 = 4$ and 
in Fig. 13 for $N_5 = 6$. 
The large peak height at $\gamma^2 = 2.0$ 
for the $8^4 \times 4$ lattice
and at $\gamma^2 = 3.6$ and $3.8$ 
for the $8^4 \times 6$ lattice (see Fig. 14)
signals
the first order transition which  we have seen in the previous subsection.
In Fig. 15, we see flip-flop in the history of the plaquette values, 
which is another sign for the first order phase transition.
The transition point $\beta_C$ and its derivative 
${\rm d}\beta_C/{\rm d}\gamma$ for a given $\gamma$ 
are given in Table II. 
Here, the derivative of a transition point is calculated
by fitting the 
continuous $\gamma$-dependence  of $ \beta_{C}$
 with the polynomial 
\be
 \beta_{C} (\gamma) &= &\sum_{n = 0}^{n_{\rm max}} f_n \, 
(\gamma - \gamma_0)^{n}~,
\label{eq:polynomial}
\ee
where $f_n$'s are  fitting parameters, and
${\rm d} \beta_{C} / {\rm d} \gamma  = f_{1}$ at $\gamma=\gamma_0$.
The range of $\gamma$ and $n_{\rm max}$ are chosen
such that the results of the 
${\rm d} \beta_{C} / {\rm d} \gamma$ are independent of 
the fitting range and the fitting function.
We adopt $\pm 0.005$ from the simulation point 
as the fitting range of $\gamma$ and 
the $n_{\rm max} = 3$ 
for the final results, respectively.
The bin size of the jackknife error analysis is 1000.

The transition points in the $\beta-\gamma$ plane are shown in Fig. 16, where
the circles $(\circ)$ are the results for $N_5 = 4 $ and the
diamonds $(\diamond)$ are those for $N_5 = 6$, respectively.
The short lines on these symbols denote the upper and lower bound 
of the slope of the transition curve. 
Two solid lines show the boundaries of the region 
in which two kind of phases coexist. 
Note that these boundary lines in Fig. 16 are obtained
in the uncompactified theory.
(Fig. 10 is an illustration of Fig. 16 transformed
into the $\beta_4-\beta_5$ plane.)
The interpolation curves are the dashed curves in Fig. 16, 
which are determined from the positions of $\beta_C(\gamma)$ 
and its slopes.  
As we see from the figure, the critical lines bend strongly 
at $\beta \approx  1.71$ and $\gamma \approx  1.42$ for 
the $8^4\times 4$ lattice, 
and $\beta \approx  1.75$ and  
$\gamma \approx  2.0$ for the $8^4\times 6$ lattice. 
The bending points are  the merging points of two transition lines, 
the one for the phase transition characterized 
by the second order transition of the transverse 
Polyakov loop (\ref{polyakov})
and the other one by the first order transition 
that is insensitive to $N_5$.

\subsection{Calculation of $\bar{\beta}_{\rm lat}$}

Using 
the data given in Table I and II, we can express $\beta$ function 
in terms of $s$,
where $s$ is given in Eq. (\ref{es}).
Then it is straightforward to compute $\bar{\beta}_{\rm lat}$
from Eq. (\ref{beta2}).
The results are shown in Fig. 17,
where the $\circ$ points are obtained
on the critical line with $N_5=4$ and the $\Box$ points 
are those with $N_5=6$.
As we see from Fig. 17, we obtain the same $\beta$-function
for two different $N_5$ (or $a_{5c}$).
This implies  that the lattice system
on two different critical lines 
describes the same physical system, and
finite $N_5$ or equivalently finite $a_5$ effects may
be regarded as negligibly small.
In Fig. 18 we show $g^{-2}$ defined in (\ref{g-def}) obtained from the data.
This data  indicate that  $g^{-2}$ depends only
on the variable $s$,
supporting our assumption that the
critical compactification radius $R_C$ is a physical quantity. 
Moreover, the Fig. 18 suggests that $g^{-2}(s)$ is almost a liner function.
Its theoretical interpretation will be given
in next subsection.
Note that the result above obtained 
for $\bar{\beta}_{\rm lat}$ does not verify
the assumption that the compactification radius $R$ is kept fixed
at the critical value $R_C$ along the line of the
phase transition due to the compactification.
To verify this assumption we need an analytical consideration 
as we will see in the next subsection.

At this point we should emphasize that, in the region with small $g^{2}$, 
the transition is of first order and is not related to the compactification.
It implies that there is no reason to assume that the compactification 
radius is $R_C$  near the first order phase transition.
In Fig. 18, the order of the transition turns to be of first order 
around $g^{-2} = 0.95$ for both cases of $N_5=4$ and $6$.
Therefore, the reliable region 
in which the compactification can be assumed to be $R_C$,
is $g^{-2} < 0.95$.
We, however, will assume in the next section, that the line of $R=R_C$ exists,  
departing  from the transition line around $g^{-2} = 0.95$ and
entering  into the deconfinement phase.
How this line extends into the deconfinement phase
cannot be found out within the framework of
the Monte-Carlo simulations; 
we need analytical considerations as we will do
in the next subsection.
There we will discuss the theoretical interpretation of our data, 
and extrapolate the line of
$R=R_C$ into the region of a smaller $g^{2}$.

\subsection{The $\epsilon$ expansion, the power-law behavior and 
the ultraviolet fixed pont}

The power-law behavior of the gauge 
coupling  is indeed suggested by its canonical 
dimension,
dim$[g] = (4-D)/2$, where $D$ is the number of
the space-time dimensions.  In 
the various explicit computations in perturbation theory
\cite{dienes1,peskin,kubo2},
this behavior has been directly seen. However, the explicit computations 
have been carried out basically
within the frame work of perturbation theory,
and so the result may not be trustful because
the theory is perturbatively 
non-renormalizable \footnote{
The result of \cite{clark,kubo2} goes slightly beyond the perturbation theory
because, though a number of 
non-trivial truncations to define an
approximation scheme should be introduced, it is based on the exact 
Wilson renormalization group approach \cite{wilson1}.}.

The simplest way to see the power law behavior 
in perturbation theory may be
in the dimensional regularization scheme,
as we do it briefly.
Let $g_D$ be the dimensionless gauge coupling in 
the pure $SU(N_C)$ Yang-Mills theory
in $D=4+\epsilon$ dimensions. The $\beta$-function is 
given by \cite{peskin,okamoto1}
\be
\beta_D &=&  \Lambda \frac{d g_D^2 }{d \Lambda} 
=\epsilon g_D^2 +\frac{2 b}{16\pi^2} g_D^4+O(\epsilon^2)~
~\mbox{with}~~~b=-\frac{11}{3}N_C~.
\label{beta3}
\ee
Now to mimic the dimensionless gauge coupling defined in
the compactified theory (see (\ref{g-def})), we introduce
\be
\tilde{g}^2 &=& \frac{g_D^2}{(2\pi R \Lambda)^{D-4}}~,
\label{g-def2}
\ee
whose $\beta$-function becomes
\be
\tilde{\beta } &=& 
\Lambda\frac{d \tilde{g}^2}{d \Lambda} =
\frac{2 b}{16\pi^2}(2\pi R\Lambda)^{D-4}~\tilde{g}^4+\dots\\
&= & -\frac{11}{12\pi^2} (2\pi R\Lambda)~\tilde{g}^4+\dots
~~\mbox{for}~~N_C=2~,~D=5~.
\label{beta4}
\ee
Eq. (\ref{beta3}) suggests that there could exist a non-trivial ultraviolet
fixed point for $g_D$, and as we have mentioned in Introduction,
this possibility in the uncompactified theory was
 ruled out by the numerical studies of
 Refs. \cite{creutz1,okamoto1,nishimura1}. 
Note that if the fixed point is real, then  it means that
the redefined coupling $\tilde{g}$ behaves as an asymptotically free
coupling,
because
\be
\tilde{g}^{-2} &\to& -\frac{2 b}{16\pi^2}
\frac{(2\pi R\Lambda)^{D-4}}{(D-4)}
\to \infty~~\mbox{as}~~\Lambda
\to \infty~.
\label{g-infty}
\ee
Translated into $g_D$, we obtain
\be
g_{D}^{-2} &=& \frac{
\tilde{g}^{-2}}{(2\pi R\Lambda)^{D-4}}
\to  -\frac{2 b}{(D-4)16\pi^2}~~\mbox{as}~~\Lambda \to \infty~,
\label{g-infty2}
\ee
which is consistent with the fixed point value obtained from
Eq. (\ref{beta3}).

The form of the lattice $\beta$-function 
 in perturbation
theory may be derived from the $\beta$-function
(\ref{beta4}), if we know the relation between $\Lambda$ and $a_4$.
Since all the (four-dimensional) momenta in a lattice theory are restricted to
the first Brillouin zone
($-(\pi /a_4) < p_\mu \leq (\pi/a_4)$), 
the momentum cut-off is $|\pi/a_4|$. That is,
$\Lambda^2 = \sum_{\mu=1}^{4} 
(\pi/a_4)^2 =(2\pi/a_4)^2$,
which implies that 
\be
\Lambda &=& \frac{2\pi}{a_4}~.
\label{cut-off}
\ee
So, the suggested one-loop lattice $\beta$-function is
\be
\beta_{\rm lat}^{(0)} &=& 
 -\frac{11}{12\pi^2} s~g^4~,
\label{one-loop1}
\ee
where \footnote{
So far there exists no perturbative computation
of $\beta_{\rm lat}$ in literature. Note also that
the one-loop coefficient of $\beta_{\rm lat}^{(0)}$
depends not only on the regularization
employed, but also on the definition of 
the gauge coupling. Our definition is given in Eq. (\ref{g-def}).}
we have used $s=2\pi N_5/\xi$,  
$R=N_5 a_5/2\pi$ and $\xi=a_4/a_5$.

\subsection{The power law from the data}

Now we would like to proceed with our numerical analysis.
Since the data in Fig. 18 suggest that $g^{-2}$ can be approximated by 
a linear function in the region we investigate, 
the one-loop form of the $\beta$-function 
(\ref{one-loop1}) is  approximately correct. 
So we fit the function of $s$ with the form 
\be
g^{-2}_p &=& C_1 + \frac{C_2}{12 \pi^2} ~s
\label{fit2}
\ee
for the data of $g^{-2}$  at the second order transition. 
We find that the best values for $N_5=4$ are
$C_1= -0.208(8)$ and $ C_2=9.84(17)$ with $\chi^2/df= 1.4$, 
and those for $N_5=6$ are
$C_1= -0.263(15)$ and $C_2=10.71(21)$ with $\chi^2/df= 0.4$, respectively.
In Fig. 18 we compare the data for $g^{-2}$ with $g^{-2}_{p}$ for $N_5=6$.
As we can see from Fig. 18, 
the one-loop ansatz (\ref{fit2}) fits well to the  data, 
and moreover, the coefficient in front of $s$
on the right-hand side of Eq. (\ref{fit2}) is close to 
the one suggested in Eq. (\ref{one-loop1}). 
Since the data with $N_5=4$ and $6$ seem to lie on the same line, 
we also fit these data simultaneously. 
We obtain $C_1= -0.224(6)$ and $ C_2=10.16(11)$  with  $\chi^2/df=1.7$, which
is a reasonable value,  implying that the fitted $g^{-2}_{p}$'s 
for different $N_5$ agree with each other.
The fact that our data have a one-loop interpretation indicate
that the assumption that the compactification radius $R$ is kept fixed
at  $R_C$ along the line of the
phase transition due to the compactification may be  correct.

Next,  to discriminate the logarithmic behavior
we would like to 
try to fit for the date on $g^{-2}$  a function of the form
\be
g^{-2}_l &=& B_1+\frac{B_2}{16\pi^2} \ln s ~,
\label{fit4}
\ee
and find that $B_1=-0.836(14)$ and $B_2=100.9(1.1)$ 
using the data for $N_5=4$ and $6$.
This fit is not a good one because we obtain $\chi^2/df=33$.
Moreover, the coefficient
$B_2$ for the logarithmic function (\ref{fit4}) cannot
be explained within the frame work of perturbation theory.
Namely, if the compactified theory on a lattice
is simply a four-dimensional theory with
Kaluza-Klein excitations of a finite number 
 $n$, then the coefficient
$B_2$ should be equal to $(40/3) n$.
Since $n$ could vary between  $1$ and $N_5=6$, 
perturbation theory for this assumption would predict
\be
13 \lsim B_2 \lsim 80~,
\label{pred3}
\ee
which clearly disagrees with the value of $B_2$ obtained from 
fitting for the data. 
Since we have found that the one-loop form of the 
power law behavior describes the data well, 
the higher order contributions,
especially those coming from non-renormalizable operators
(remember the naive continuum theory is not renormalizable by
power counting) must be suppressed, at least in the parameter
region in our numerical simulations.

It is the subject of the next section to
investigate this possibility,
where we will assume that the theoretical function (\ref{fit2}) can be used
to draw the lines of $R=R_C$ even
in the deconfining phase of the transverse Polyakov loop (\ref{polyakov}).

\section{TOWARD A CONTINUUM LIMIT}

In the weak coupling  regime, which is the most important regime
to investigate a continuum limit,  the way
to use the transverse string tension as a physical
quantity to give a physical scale
 is not available,  because the phase transition 
due to the compactification in that regime disappears.
One of the central assumption in discussing the continuum limit
in this paper is that
the one-loop function (\ref{fit2}) can 
be extended into the weak coupling regime.
Equivalently,
we assume that the $\beta$-function for a given
compactification radius $R$ is given
by
\be
\beta_{\rm lat} &=&-a_4\frac{d g^2}{d a_4}=
-\frac{C_2}{3 }~\frac{R}{a_4}~g^4~,
\label{beta-th}
\ee
where $C_2$ is given in Eq. (\ref{fit2}).
The assumption implies that we can draw lines of
$R=$ const. in the weak coupling regime.
On these lines $\beta$ becomes a function of $\gamma$,
which can be obtained from
TABLE I, (\ref{g-def}) and (\ref{fit2}).
So, given the lattice 
$\beta$-function (\ref{beta-th}) we now know
how to approach continuum limits.
The object whose scaling property should be investigated is
the longitudinal lattice string
tension $\sigma_{\rm lat}$ which we replace
by the Creutz ratio (\ref{creutz}), because, as
we have seen in section IV, the longitudinal Creutz ratio
can be non-vanishing even in the deconfining phase
of the transverse Polyakov loop.
In the following subsections we will investigate
 the  scaling law of the longitudinal string tension 
\begin{equation}
 \sigma_{\rm lat} = \sigma_{\rm phys} a_4^2 
 = \sigma_{\rm phys} a_5^2 \xi^2~,
\label{eq:sigma1}
\end{equation}
where we have assumed that  
$\sigma_{\rm phys}$ should remain finite in the continuum limit. 

\subsection{$a_5 \to 0$ limit}
We apply the  scaling hypothesis (\ref{eq:sigma1}) 
to the $a_5 \to 0$ limit with
$\xi$ fixed at a certain value.
As we have stated,
we assume that 
the one-loop function (\ref{fit2}) can 
be extended into the weak coupling regime.
If we move along the line of $\xi$=const., we change the
compactification radius $R$.
To express this more precisely, 
we first derive the  scaling law for this case.
To this end, let us consider the lines of $R=$ const.
for various $N_5$ 
in the parameter space $(\beta, \xi)$,
where it is implicitly assumed that the constants
$C_1$ and $C_2$ in Eq. (\ref{fit2}) are independent of $N_5$
(the consistency of this assumption is checked 
for $N_5= 4$ and $6$ in section V):
\be
\beta &=& -\frac{D_1}{N_5}\xi+D_2~,
~D_1=-4 C_1~,~D_2=\frac{2 C_2}{3 \pi}~~.
\label{beta-on}
\ee
Since 
$a_5 = R_C/(2 \pi N_5)$ and 
$R_C$ is assumed to be a physical quantity, 
we obtain from (\ref{beta-on})
\be
a_5 &\propto & (N_5)^{-1} \propto (\beta-D_2)~.
\label{a5}
\ee
 Inserting Eq. (\ref{a5}) into
(\ref{eq:sigma1}), we find that
\be
\ln \sigma_{\rm lat} &=&2\ln |\beta-D_2|+ ~\mbox{const.}
\label{sigma3}
\ee

Here, we use the value of $D_2$ obtained by fitting the data 
for $N_5=4$ and $6$ simultaneously.
The result of this scaling behavior 
is shown in Fig. 19, where we have used the $8^4\times 4$
lattice with $\gamma=2.0$ as for Fig. 11. 
The bold line corresponds to the theoretical
line (\ref{sigma3}). We see from Fig. 19 that the scaling law 
 (\ref{sigma3}) is well satisfied for $ 1.6 \lsim \beta \lsim 1.8$.
Below $\sim 1.5$ we enter into the region
of the strong coupling, and, above $\sim 1.9$, 
finite size effects due to small $a_4$
presumably start to become visible.
So we may conclude that the data are consistent with the scaling law
 (\ref{sigma3}).  This is an important result,   and is indeed
the only result
which supports the correctness of the assumption that 
the lattice spacing has a physical scale even 
in the deconfining phase
of the transverse Polyakov loop and of our way how
to extend the lines of $R=$ const. into that phase:
This is an evidence for the existence of the fixed point
suggested in the $\epsilon$ expansion.

\subsection{$a_4/R \to 0$ limit}
Next, we would like to investigate the scaling behavior 
of the longitudinal lattice string
tension 
in the $a_4/R \to 0$ limit with $R$ kept fixed.
Since $2\pi R=N_5 a_5$, the lattice spacing $a_5$
is kept fixed in this limit for a given $N_5$.
Then, the string tension should obey the  scaling law,
\begin{equation}
 \sigma_{\rm lat} 
 \propto \xi^2 ~~\mbox{or}~~
\ln  \sigma_{\rm lat}=2 \ln \xi + ~\mbox{const.}~.
\label{eq:sigma}
\end{equation}

We compute on $8^4 \times N_5$ lattices with 
$N_5= 2,3,4,5,6$ and $8$ 
the longitudinal Creutz ratio $\chi (i,j)$ 
along the theoretical line of $R=$ const. 
on which $N_5=6$ is critical.
To determine this line, we used 
the $C_1$ and $C_2$ in Eq.(\ref{fit2}) obtained from the data of $N_5=6$.
Note that for a given $N_5$  the compactification radius $R$ 
is $N_5 a_5/(2\pi) = (N_5/6) R_C$.
In Figs. 20, we plot $\ln \chi (i,j)$ as a function of $\ln \xi$.
If the slope of the $\ln \chi (i,j)$ 
is equal to $2$, the scaling relation of Eq. (\ref{eq:sigma}) is 
realized. 
In the  $N_5 = 8$ case, the results of the ordered start 
and disordered start are split, and
the  longitudinal  Creutz ratio $\chi(i,j)$ 
with large $i$ and $j$ of the ordered start 
fall drastically
when we move from a large $\xi$ 
to a small $\xi$. 
This is in accord with our expectation,
because the lattice system corresponds to the uncompactified.
In the $N_5 =6$ case 
 (the compactification radius $R$ is equal to $R_C$)
  the  longitudinal  Creutz ratios also
start to fall down around $\ln\xi\approx 1.1$.
Therefore, the lattice system above does not correspond to
any four-dimensional theory,
rather it describes a full five-dimensional theory.
Keeping this in mind,
we continue to consider the $N_5=5, 4, 3$ and $2$ cases.

The results are also shown in Fig. 20.
As we see from these figures, the  longitudinal Creutz ratios 
no longer  fall drastically.
Comparing
the slope of these  longitudinal Creutz ratios with the 
straight lines of the slope 2, we find that
the  longitudinal Creutz ratios for $N_5 \gsim 3$ 
decrease faster than $\xi^2$ as $\xi$  decreases.
 If this continues to smaller $\xi$,
i.e. smaller $a_4$, we may conclude that
that for $N_5 \gsim 3$  
 the string tension $\sigma_{\rm phys}$ decreases as $a_4$ decreases so that
$\sigma_{\rm phys}$ vanishes in the $a_4/R \to 0$ limit.

As we have observed, the slope of the Creutz ratios becomes milder
as $R$  decreases.
This tendency of the milder-becoming slope with decreasing
$R$ is a real effect and 
not an effect of a finite $N_5$, at least $N_5 \geq 4$
or equivalently $R  \geq (2/3) R_C$, which we may conclude
from the fact that our data show
that the critical lattice systems
with $N_5=6$ and $4$ describe the same physical system.
Some finite size effects may be present in the case of
$N_5=3$  and $2$ in Fig. 20. Nevertheless,  the tendency can be seen
for these cases, too.
It is  this tendency of the milder-becoming slope 
that suggests the existence of a continuum 
theory with a non-vanishing string tension.
If we assume that in the present case of setting
the longitudinal Creutz ratio
starts to scale according to the scaling law (\ref{eq:sigma})
from  $N_5=2$ on, we obtain the maximal compactification
radius 
\be
R_M &\approx & \frac{R_C}{3} 
\approx \frac{1}{10 \sqrt{\sigma_{\rm phys }}}~,
\ee
below  which the compactified theory
with a non-vanishing string tension could exist non-perturbatively.

We would like to notice that,
though the qualitative nature
of the milder-becoming slope is real,
the  scaling behavior of the
longitudinal Creutz ratio itself
is sensitive  to the choice of 
the extrapolation function (\ref{fit2}) that describes
the lines of  $R=$ constant.
It is therefore clear that for a more definite 
conclusion more refined analyses with
a lager size of lattice are indispensable.

\subsection{``Simulated'' $N_5 \to \infty$ limit}

To consider the $N_5 \to \infty$ limit with
$R=$ const., we  have 
to enlarge the size of our lattice. Instead of enlarging the size, however,
we can simulate the limit
with the data that we have already at hand.
We would like to argue below that 
 the second limiting process,
 $a_4/R \to 0$ with $a_{5}$ fixed at an $a_{5c}$
(see Fig. 19) 
can be interpreted as
an $a_4, a_5 \to 0$ limiting process with $R$ fixed.
($a_5 \to 0$ with $R$ fixed is the same as 
$N_5 \to \infty$ with $R$ fixed.)

We have been assuming that
the theoretical function given in
Eq. (\ref{beta-on})  describes a set of
the lines of $R=$ const. in the $\beta-\xi$ plane 
for different  $N_5$.
All  lines so obtained are assumed to be physically equivalent:
To each point on a line, there exists an equivalent
point on each line.
It follows then that all the points on a line 
described by Eq. (\ref{beta-on}) for a given  $N_5$  can be transformed
 into a line that is parallel to the $\beta$
axis in the $\beta-\xi$ plane. The mapping can be easily
found, because the values of the gauge coupling $g$
on the physically equivalent points should be the same.
Since $\beta$ does not change if the ratio
$\xi/N_5$ is fixed (see Eq. (\ref{beta-on})),
the value of $g$ does not change if we move along
a line parallel to the $\xi$ axis (see Eq. (\ref{g-def})).
That is, to find a set of physically equivalent points we just
have to move parallel to the $\xi$ axis.
Therefore, moving along a line described by Eq. (\ref{beta-on}) 
for a given  $N_5$ can be assumed to be
physically equivalent to moving along a line with $\xi=$ const.
while changing $\beta$ and $N_5$: 
We can simulate enlarging $N_5$ without changing $N_5$.
Since the compactification radius $R$ is assumed to
be  a physical quantity,
it remains unchanged during the transformation.

Consequently, the scaling behavior of the Creutz ratios studied 
in Fig. 20 can be reinterpreted as the scaling behavior 
along a line with $\xi$ and $R$ kept fixed, where
the scaling law appropriate for this limiting process is given
in Eq. (\ref{sigma3}):
The vertical axis $\ln \xi$ in Fig. 20 should be replaced by
$\ln |\beta-D_2|-0.97$ and the straight line should
be understood as 
$\ln \xi = 2 \ln| \beta -D_2| +$ const.,
where we have used Eq. (\ref{beta-on}).
We arrive at the same
conclusion 
as in the  $a_4 /R \to 0$ case,
which we do not repeat here again.
But as we have stated there, the tendency of 
the milder-becoming slope with decreasing
$R$ is a real effect, at least for $R  \geq (2/3) R_C$.
This is so here, too,  because our data show
that the critical lattice systems
with $N_5=4$ and $6$ describe the same physical system
so that the above mentioned
transformation at least 
between the $N_5=4$ and $6$ lines
is trustful.
The simulated $N_5  \to \infty$ limit we have
considered here should be regarded as a prediction
of the real limit, at least  for $R  \geq (2/3) R_C$.

\section{SUMMARY AND CONCLUSION}

Our motivation in this paper has been to see,
within the frame work of the lattice gauge theory, whether or not
the non-trivial fixed point found in the $\epsilon$-expansion
in the continuum theory of the pure $SU(2)$ Yang-Mills theory
in five dimensions
is spurious in the case that the fifth dimension is
compactified. 
We have used intensively anisotropic lattices
to take into account the
compactification. We have found that
the compactification changes the nature of the phase transition:
A second order phase transition,
which does not exist in the uncompactified case,
begins to occur, and
turns to be of first order at a certain point.

Under the assumption that
the compactification
radius  $R$ remains  constant fixed at the critical value
$R_C$ along the critical
lines of the phase transition due to the compactification,
we have computed the lattice
$\beta$-function $\bar{\beta}_{\rm lat}$, and found that
$\bar{\beta}_{\rm lat}$ as a function
of $s$,  obtained from the critical
line of $N_5=4$ and $6$,  is the same  (see Fig. 17).
We have also found that the gauge coupling
on  these critical
lines is the same (see Fig. 18).
From these observations we have concluded that the critical
lattice system  with $N_5=4$ and $6$ describes the same physical
system, and we are led to the assumption that
this is the case for all $N_5$.

As we can see from Fig. 18, 
the power-law running of the gauge
coupling (the solid line) is consistent with the data,
which has a simple one-loop interpretation.
This is the fact that supports the correctness
of the assumption, at least for $N_5=4$ and $6$,  that
the compactification
radius $R$ remains  constant fixed  at the critical value
$R_C$ along the critical
lines of the phase transition due to the copmpactification.

At this point it is the natural thing to extend
our findings into the deconfining phase of the transverse Polyakov loop:
We have assumed that
the lattice spacing has a physical scale even 
in the deconfining phase and 
the one-loop ansatz (\ref{fit2}) can be used to
draw the lines of $R=$ const. in that  regime.
The investigation of the scaling law (\ref{sigma3}) 
for the longitudinal Creutz ratio  (\ref{creutz})
shown in Fig. 19 supports the correctness of this assumption.
At this stage, the existence of the non-trivial fixed point
suggested in the $\epsilon$ expansion might be
evident.

We have investigated the scaling behavior 
of the longitudinal Creutz ratio
in the $a_4/R \to 0$ limit with $R$ kept fixed, and
found that the slope with which the Creutz ratios 
fall in the $a_4/R \to 0$ limit becomes milder
as $R$  decreases (see Fig. 20).
From  this tendency of the milder-becoming slope 
we are led to the interpretation 
that the compactified theory 
having a non-vanishing string tension could exist non-perturbatively
if  the compactification radius $R$ is smaller than 
the maximal compactification radius $R_M$.
Our estimate is:
 $ R_M \approx R_C/3 \approx 0.1/\sqrt{\sigma_{\rm phys}}$.

It is clear that to make  our interpretation more 
solid, we need not only refined and detailed numerical analyses
but also analytical investigations.
We hope that further studies will clarify the
problems on 
the quantum  realization of the old Kaluza-Klein idea.

\acknowledgments

This work is supported by the Grants-in-Aid
for Scientific Research  from 
 the Japan Society for the Promotion of Science (JSPS) (No. 11640266).
One of us (SE) is  a JSPS research fellow.
We would like to thank for useful discussions
T. Izubuchi,  K. Kanaya, H. Nakano, H. So,  T. Suzuki and H. Terao.

\begin{table}[tb]
\caption{$\xi - \gamma$ relation}
\label{tab:xi-gamma}
\begin{center}
\begin{tabular}{cccc}
\hline
$\gamma^2$ & $\xi$ & $\beta$-range & lattice size \\
\hline
 1.50  & 1.438(57) & 1.51868 - 1.66565 & $8^4\times16$ \\
 2.00  & 1.784(50) & 1.55563 - 1.69706 & $8^4\times16$ \\
 3.00  & 2.340(40) & 1.59349 - 1.73205 & $8^4\times16$ \\
 4.00  & 2.779(34) & 1.60000 - 1.75000 & $8^4\times16$ \\
 5.00  & 3.161(39) & 1.65469 - 1.74413 & $8^4\times16$ \\
 6.00  & 3.490(33) & 1.61666 - 1.76363 & $8^4\times20$ \\
 8.00  & 4.062(39) & 1.62635 - 1.76777 & $8^4\times24$ \\
10.00  & 4.617(35) & 1.50208 - 1.73925 & $8^4\times24$ \\
16.00  & 5.923(51) & 1.50000 - 1.70000 & $8^4\times32$ \\
\hline
\end{tabular}
\end{center}
\end{table}

\begin{table}[tb]
\caption{Results for $\beta_C$ and ${\rm d}\beta_C/{\rm d}\gamma$ 
by the histogram method. The simulations
are performed at  $(\beta_{4},\beta_{5})$.}
\label{tab:betac}
\begin{center}
\begin{tabular}{cccccc}
\hline
lattice & $\gamma^2$ & $(\beta_{4},\beta_{5})$ & 
$\beta_C$ & ${\rm d}\beta_C/{\rm d}\gamma$ \\
\hline
$8^4\times4$ 
&  2.0 & (1.21250, 2.42500) & 1.71472(6)  &  $-$0.0087(58) \\
&  2.1 & (1.18350, 2.48535) & 1.71342(25) &  $-$0.067(16)  \\
&  2.5 & (1.07080, 2.67700) & 1.69060(36) &  $-$0.2219(82) \\
&  3.0 & (0.95000, 2.85000) & 1.64702(31) &  $-$0.3337(67) \\
&  4.0 & (0.77000, 3.08000) & 1.54018(48) &  $-$0.4049(79) \\
&  6.0 & (0.55100, 3.30600) & 1.34560(59) &  $-$0.426(17)  \\
&  8.0 & (0.42500, 3.40000) & 1.19790(47) &  $-$0.368(11)  \\
& 16.0 & (0.21875, 3.50000) & 0.87265(37) &  $-$0.2070(36) \\
\hline
$8^4\times6$ 
&  3.6 & (0.92750, 3.33900) & 1.75943(8)  &     0.0059(12) \\
&  3.8 & (0.90150, 3.42570) & 1.75654(19) &  $-$0.0536(71) \\
&  4.0 & (0.87750, 3.51000) & 1.75339(44) &  $-$0.088(36)  \\
&  5.0 & (0.76900, 3.84500) & 1.72102(26) &  $-$0.1723(61) \\
&  6.0 & (0.68500, 4.11000) & 1.67534(51) &  $-$0.2505(89) \\
&  8.0 & (0.55550, 4.44400) & 1.57140(61) &  $-$0.3080(82) \\
& 10.0 & (0.46400, 4.64000) & 1.46861(61) &  $-$0.3194(99) \\
& 16.0 & (0.30625, 4.90000) & 1.229223(65) &  $-$0.2533(96) \\
\hline
\end{tabular}
\end{center}
\end{table}

\begin{figure}


\centerline{
\epsfxsize=12cm\epsfbox{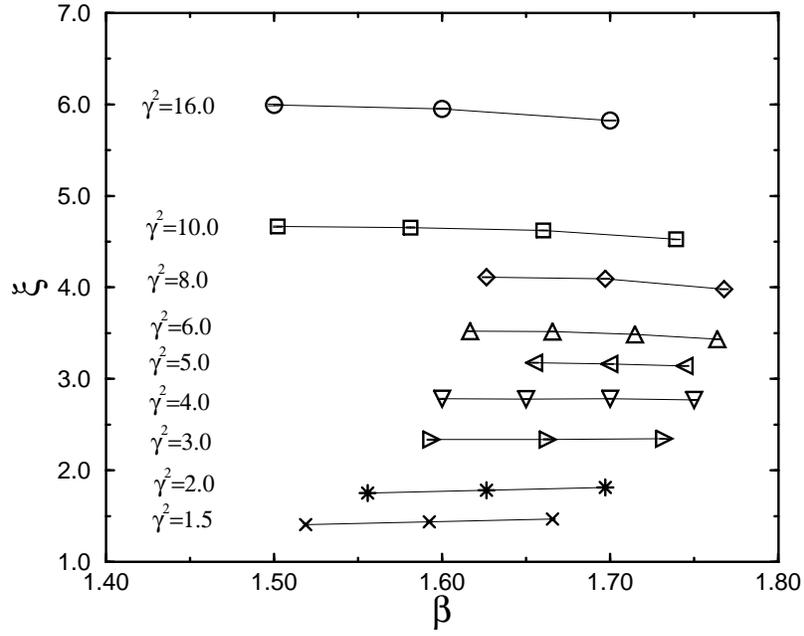}
}


\caption{
$\beta$ dependence of the anisotropy parameter $\xi$.
}

\label{fig:xi-beta}


\end{figure}

\begin{figure}


\centerline{
\epsfxsize=12cm\epsfbox{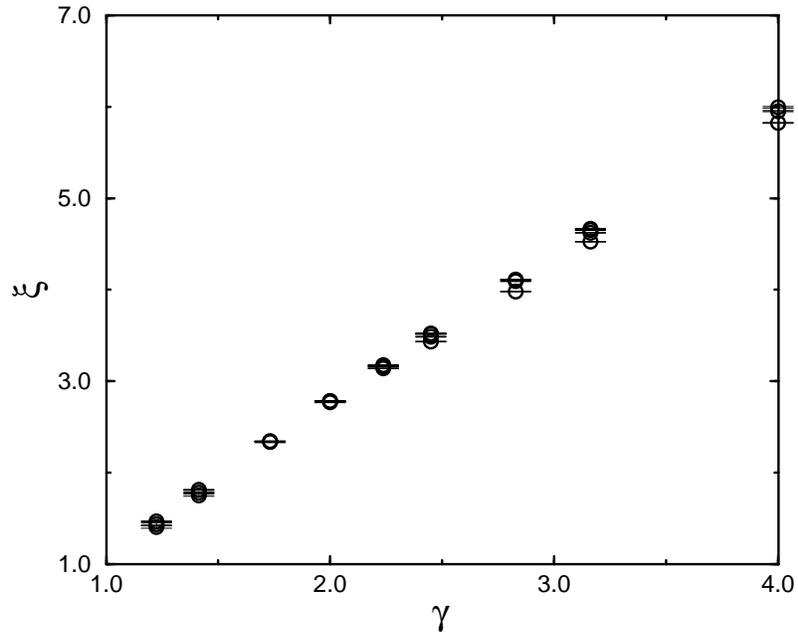}
}


\caption{
Relation between $\xi$ and $\gamma$.
}

\label{fig:xi-gamma}


\end{figure}

\begin{figure}


\centerline{
\epsfxsize=12cm\epsfbox{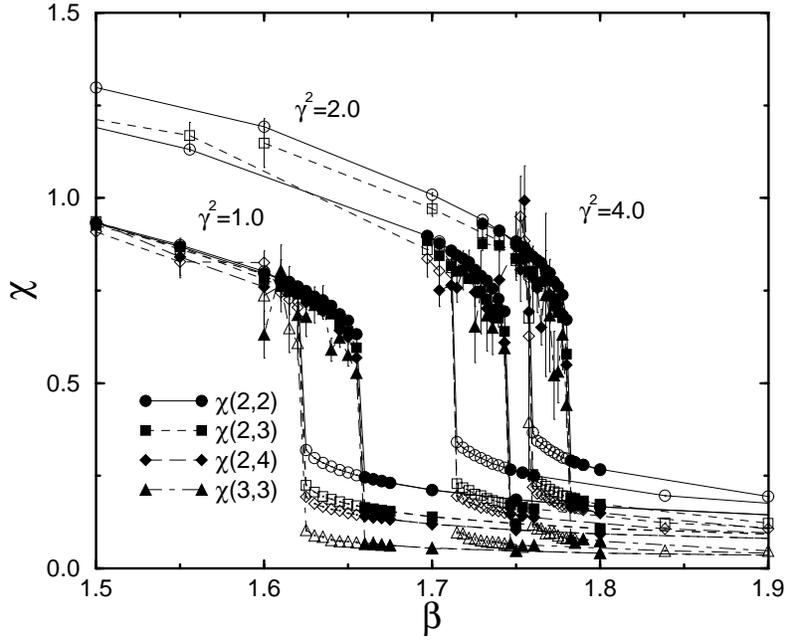}
}


\caption{
Creutz ratios as a function of $\beta$ for $\gamma^2 =1.0, 2.0$
and $4.0$ on an $8^5$ lattice.
Open symbols are the results of the ordered start and 
filled symbols are those of the disordered start.
}

\label{fig:creutz88}


\end{figure}

\begin{figure}


\centerline{
\epsfxsize=12cm\epsfbox{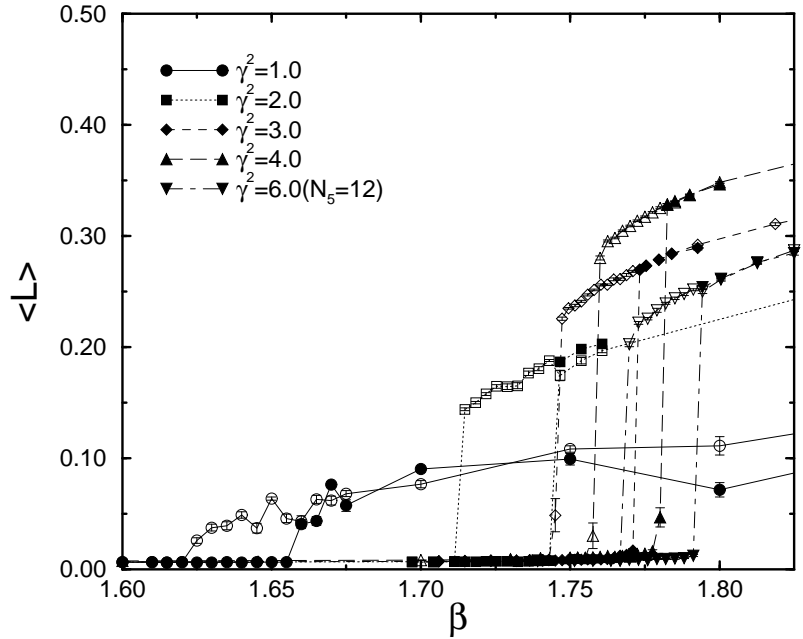}
}


\caption{
Expectation values of the transverse Polyakov loop on an $8^5$ lattice 
for $\gamma^2 = 1.0,2.0,3.0$ and $4.0$, and those on an $8^4 \times 12$
lattice for $\gamma^2 = 6.0$.
}

\label{fig:pl588}


\end{figure}

\begin{figure}


\centerline{
\epsfxsize=12cm\epsfbox{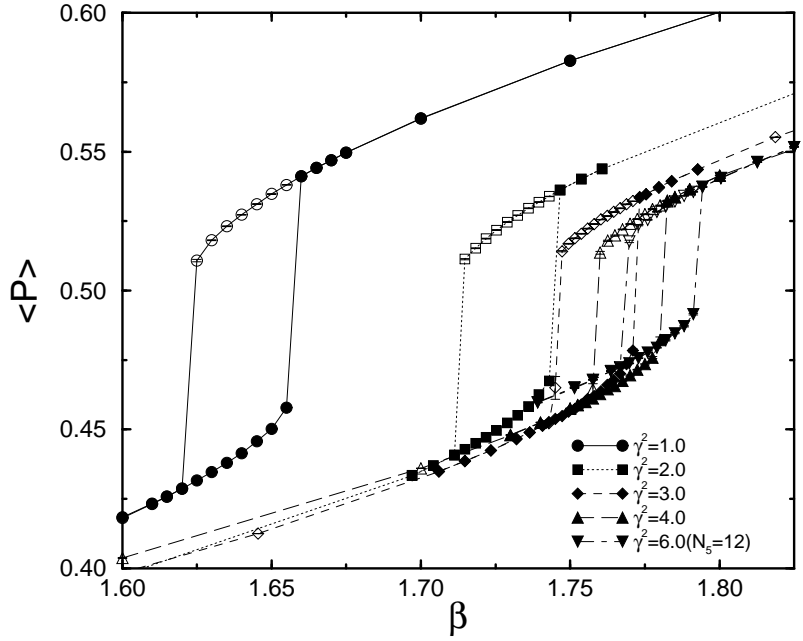}
}


\caption{
Expectation values of the plaquette on an $8^5$ lattice 
for $\gamma^2 = 1.0,2.0,3.0$ and $4.0$, and those on an $8^4 \times 12$
lattice for $\gamma^2 = 6.0$.
}

\label{fig:plq88}


\end{figure}

\begin{figure}


\centerline{
\epsfxsize=12cm\epsfbox{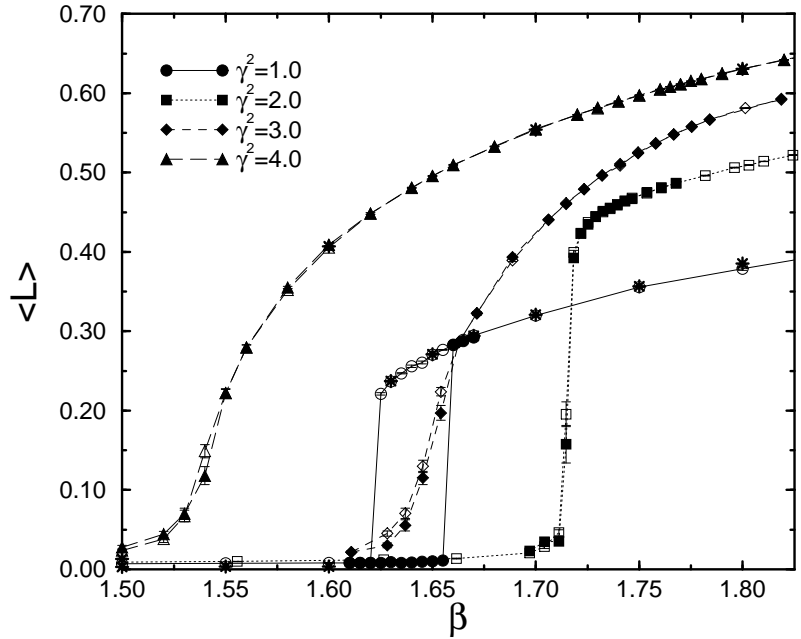}
}


\caption{
Expectation values of the transverse Polyakov loop 
on an $8^4 \times 4$ lattice. 
The star symbols are the results on a $12^4 \times 4$ lattice.
}

\label{fig:pl584}


\end{figure}

\begin{figure}


\centerline{
\epsfxsize=12cm\epsfbox{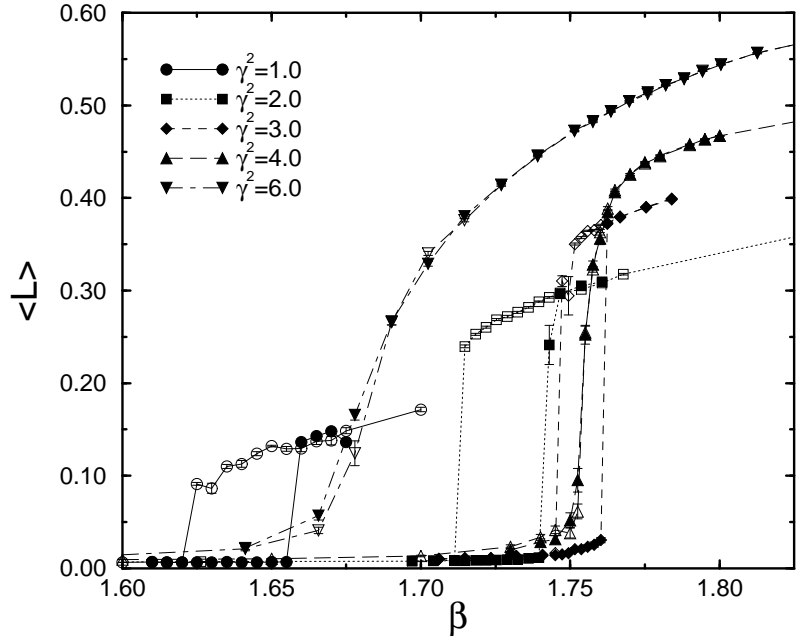}
}


\caption{
Expectation values of the  transverse Polyakov loop 
on an $8^4 \times 6$ lattice. 
}

\label{fig:pl586}


\end{figure}

\begin{figure}


\centerline{
\epsfxsize=12cm\epsfbox{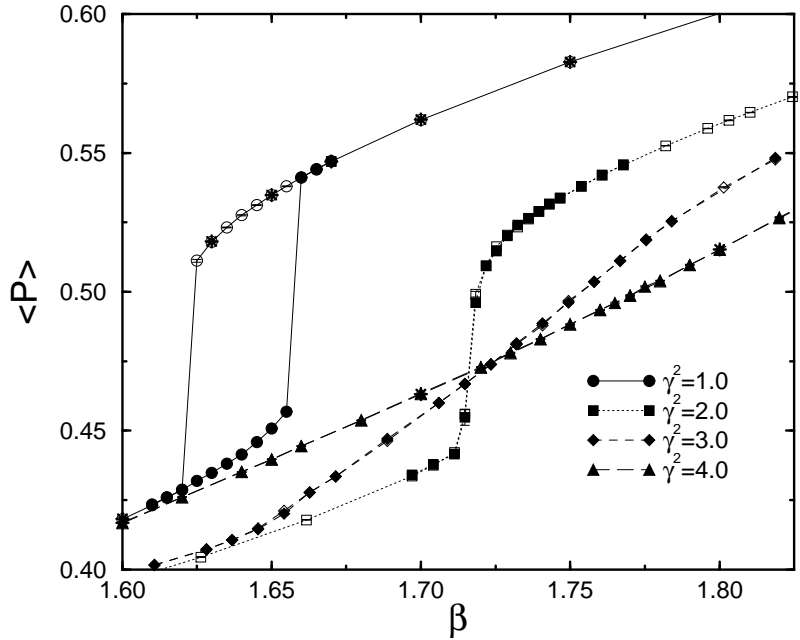}
}

\caption{
Expectation values of the plaquette on an $8^4 \times 4$ lattice. 
The star symbols are the results on a $12^4 \times 4$ lattice.
}

\label{fig:plq84}


\end{figure}

\begin{figure}


\centerline{
\epsfxsize=12cm\epsfbox{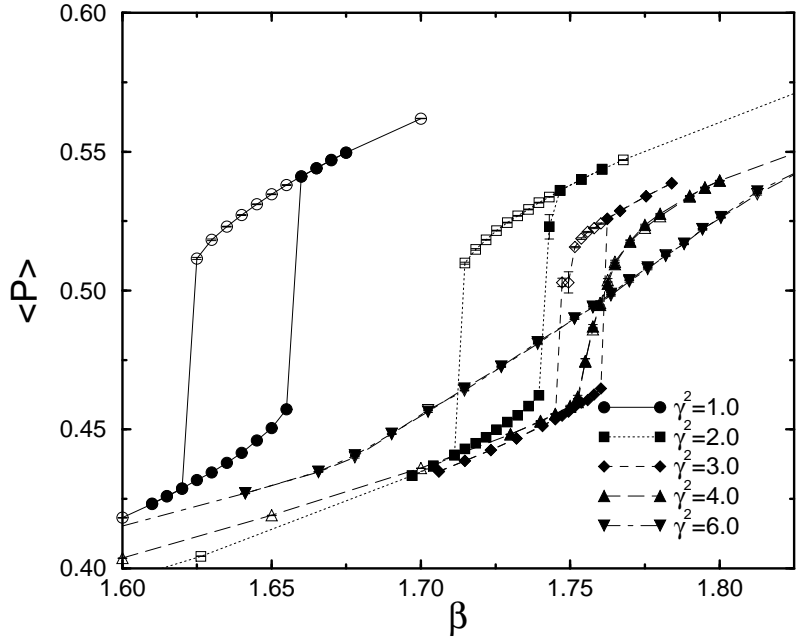}
}


\caption{
Expectation values of the plaquette on an $8^4 \times 6$ lattice. 
}

\label{fig:plq86}


\end{figure}

\begin{figure}



\centerline{
\epsfxsize=14cm\epsfbox{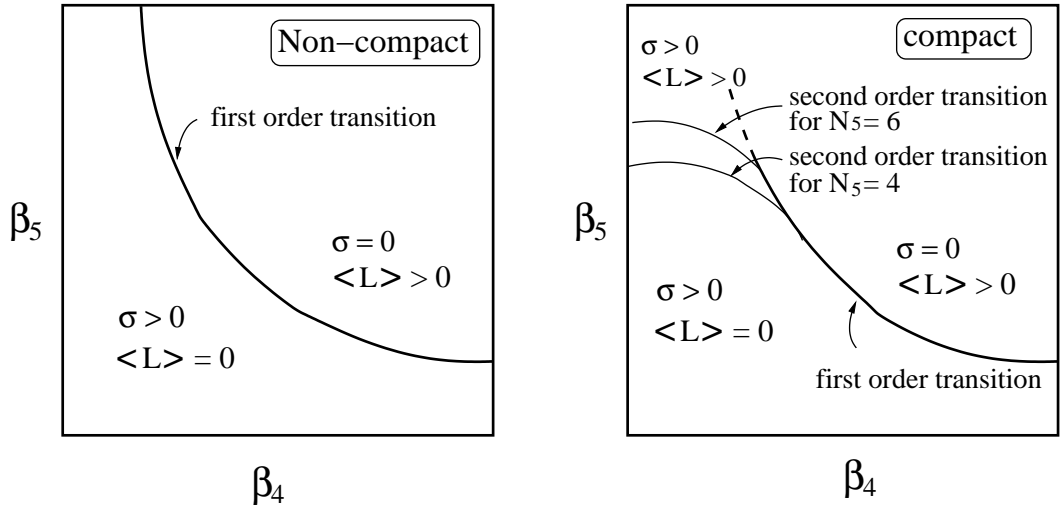}
}


\caption{
Illustrations of the phase structure 
for the non-compactified case (left) and 
the compactified case (right),
where $\sigma$ is the longitudinal Creutz ratio and
$L$ is the transverse Polyakov loop.}

\label{fig:phase}


\end{figure}

\begin{figure}


\centerline{
\epsfxsize=12cm\epsfbox{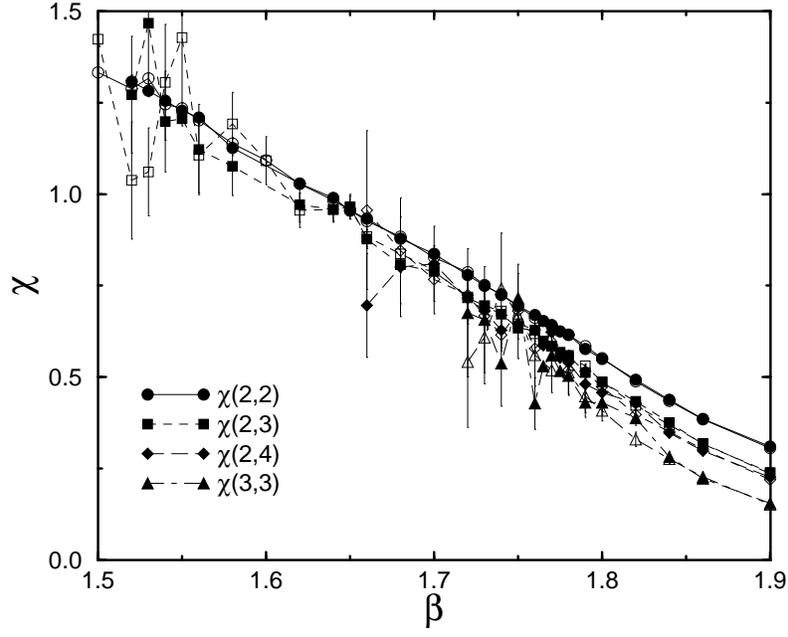}
}


\caption{
Creutz ratios on an $8^4 \times 4$ lattice at $\gamma^2 = 4.0$. 
}

\label{fig:creutz84}


\end{figure}

\begin{figure}


\centerline{
\epsfxsize=12cm\epsfbox{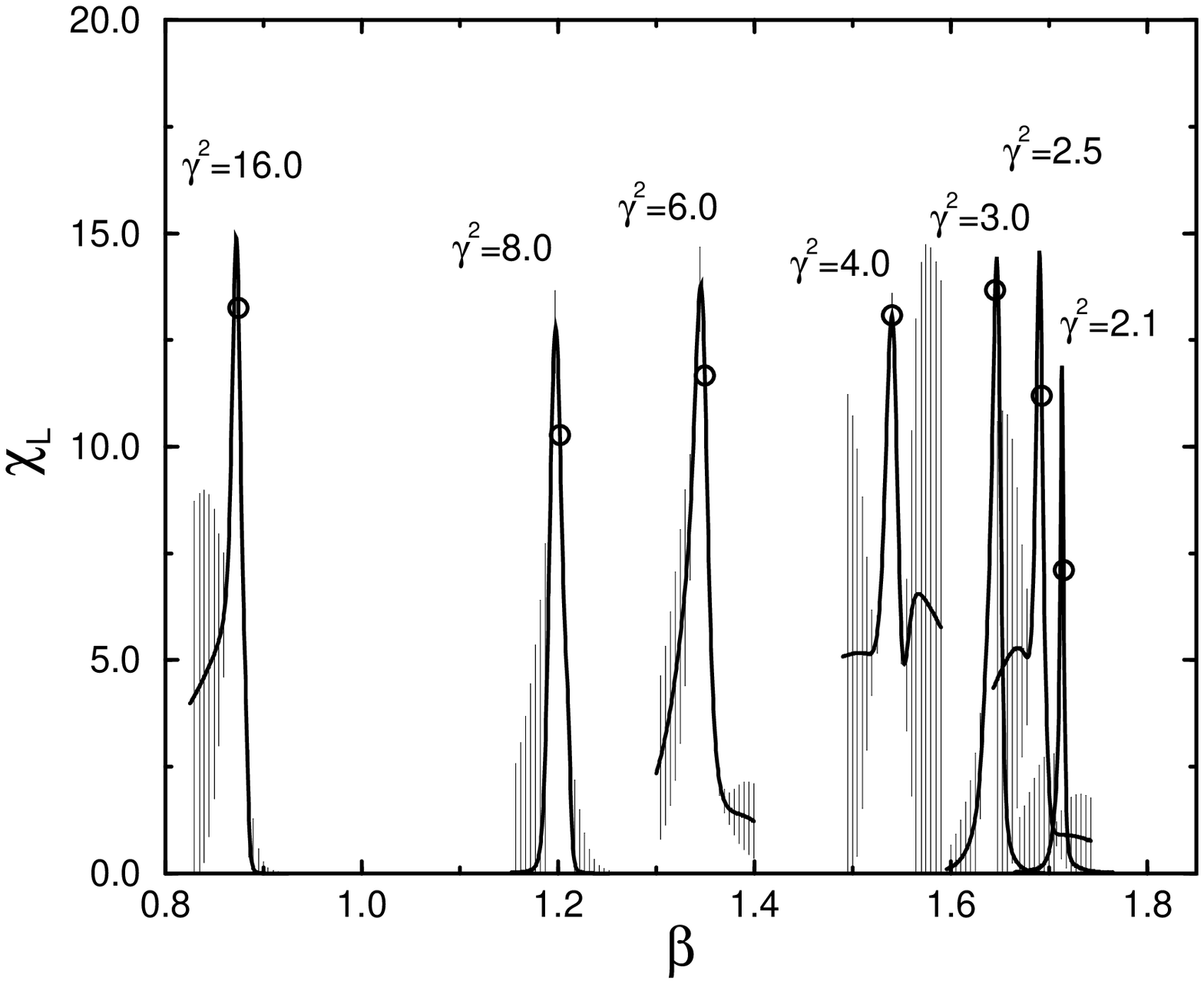}
}
\caption{$\beta$ dependence of the Polyakov loop susceptibility 
obtained by the histogram method
on an $8^4 \times 4$ lattice with $\gamma^2 \geq 2.1$.
The circles denote the simulation point.
}
\label{fig:suscept84}
\end{figure}

\begin{figure}


\centerline{
\epsfxsize=12cm\epsfbox{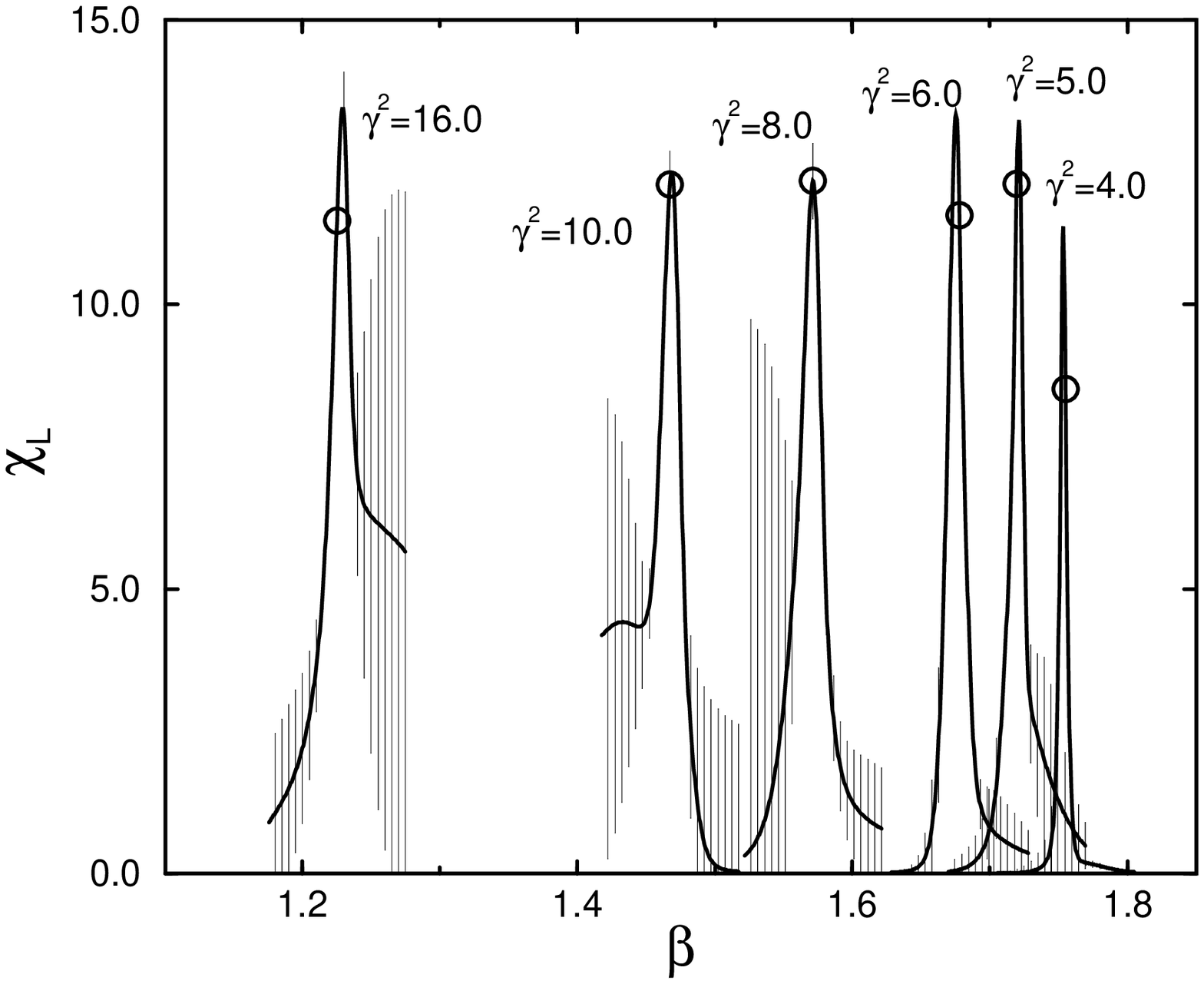}
}


\caption{
$\beta$ dependence of the Polyakov loop susceptibility 
obtained by the histogram method
on an $8^4 \times 6$ lattice with $\gamma^2 \geq 4.0$.
The circles denote the simulation point.
}

\label{fig:suscept86}


\end{figure}

\begin{figure}


\centerline{
\epsfxsize=12cm\epsfbox{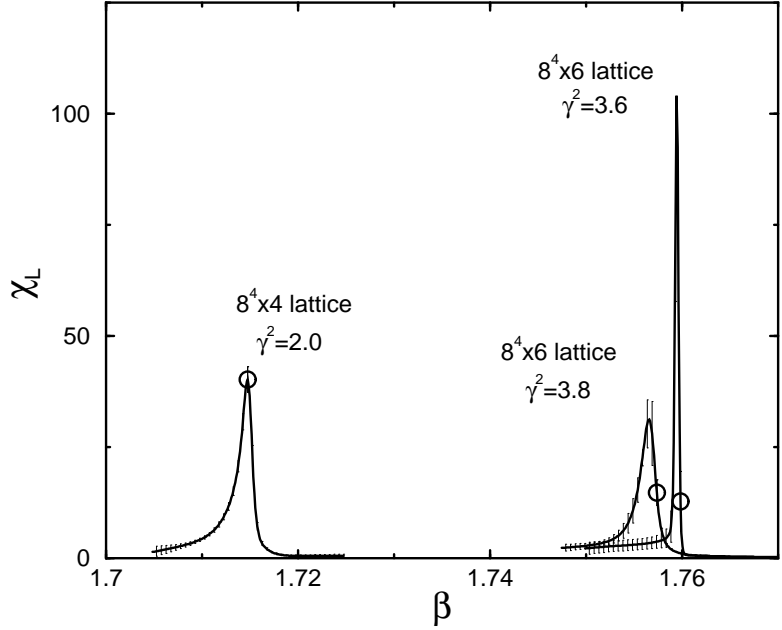}
}


\caption{
Large peaks of the Polyakov loop susceptibility 
obtained by the histogram method at $\gamma^2 = 2.0$
on an $8^4 \times 4$ lattice, and $\gamma^2 = 3.6$ and 
$3.8$ on an $8^4 \times 6$ lattice, respectively.
}

\label{fig:suscept1st}


\end{figure}

\begin{figure}

\vspace*{-5mm}

\centerline{
\epsfxsize=12cm\epsfbox{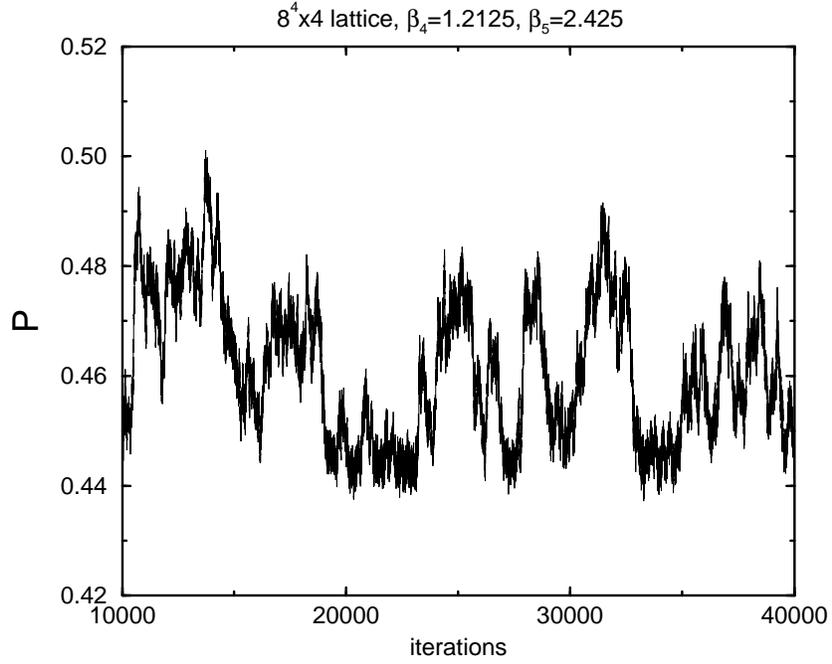}
}

\vspace*{-2mm}

\caption{
Flip-flop in the history of the plaquette value at $\gamma^2 = 4.0$ 
on an $8^4 \times 4$ lattice.
}

\label{fig:plqhis}


\end{figure}
\begin{figure}


\centerline{
\epsfxsize=12cm\epsfbox{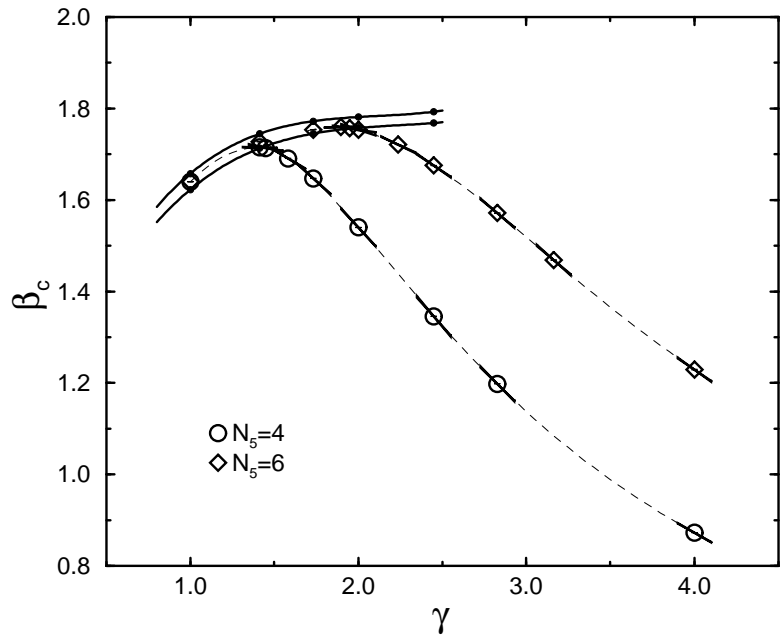}
}

\vspace*{-2mm}

\caption{
Phase transition points for $N_5 = 4$ $(\circ)$ and 
$N_5 = 6$ $(\Diamond)$ in the $\beta-\gamma$ plane.
Two solid lines denote the boundaries of the region 
in which two kind of phases coexist. 
Compare the figure with Fig. 10.}

\label{fig:betac}


\end{figure}

\begin{figure}


\centerline{
\epsfxsize=12cm\epsfbox{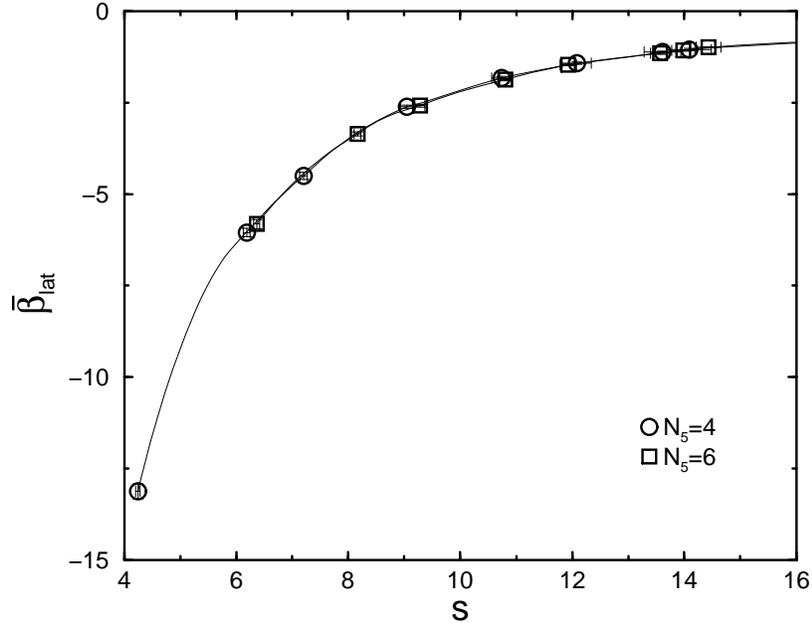}
}


\caption{
$\bar{\beta}_{\rm lat}$ as a function of $s$ determined on the 
transition lines of $N_5 = 4$ $(\circ)$ and $N_5 = 6$ $(\Box)$.
The figures shows the physical equivalence between the
critical lattice systems.}

\label{fig:betafnc}


\end{figure}

\begin{figure}


\centerline{
\epsfxsize=12cm\epsfbox{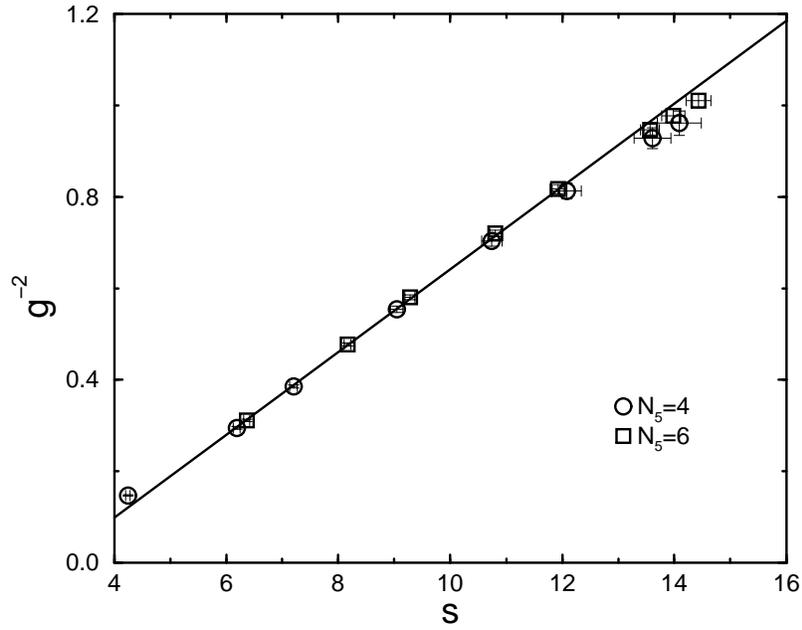}
}


\caption{The power-law behavior of
$g^{-2}$ as a function of $s$ determined on the 
transition lines of $N_5 = 4$ $(\circ)$ and $N_5 = 6$ $(\Box)$,
where the straight line is the one-loop line (\ref{fit2}).}

\label{fig:betafnc2}


\end{figure}

\begin{figure}


\centerline{
\epsfxsize=12cm\epsfbox{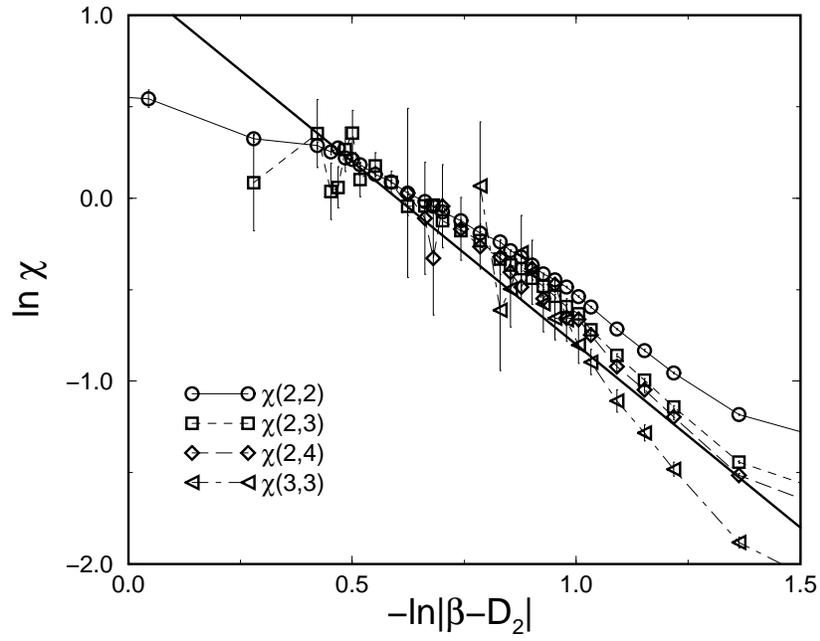}
}


\caption{
Scaling behavior of the Creutz ratio along the $\gamma^2=4.0$ line 
on an $8^4 \times 4$ lattice.
The solid line is
$\ln (\chi) = 2 \ln |\beta -D_2| +1.2$ (see Eq.(\ref{sigma3})),
where $D_2=2.156(23)$.
}

\label{fig:creuzscl2}


\end{figure}

\newpage

\begin{figure}



\centerline{
\epsfxsize=8cm\epsfbox{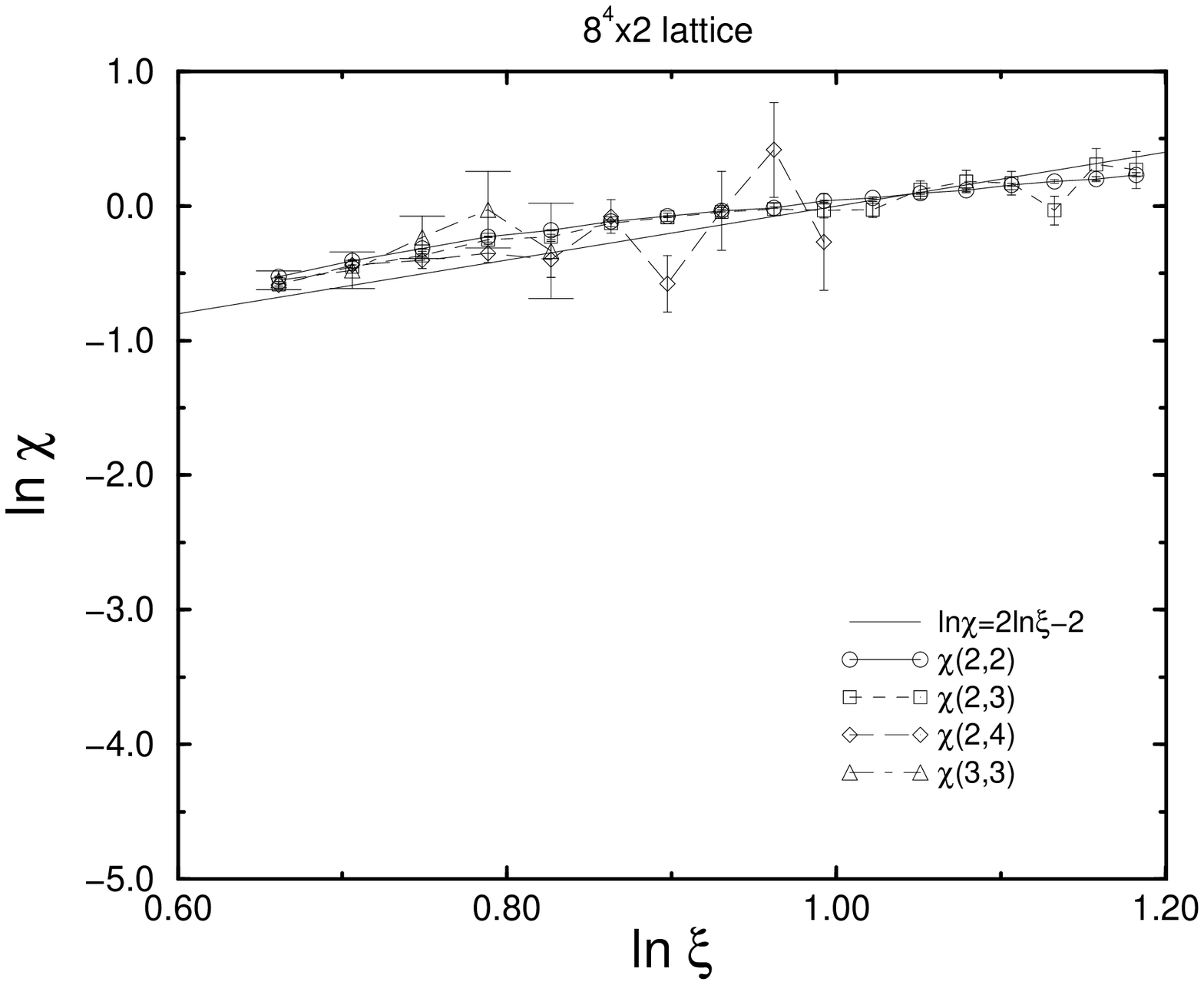}
\epsfxsize=8cm\epsfbox{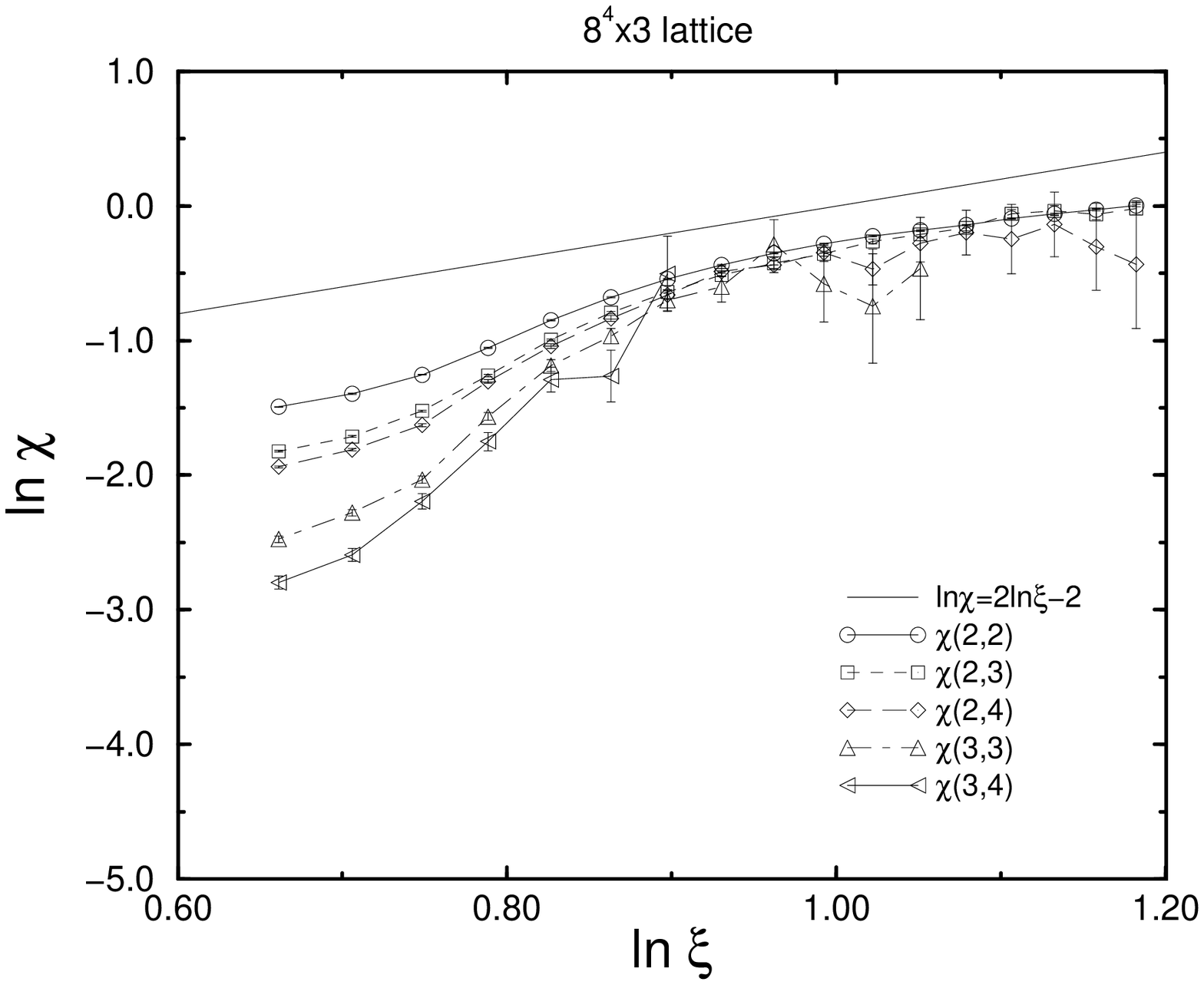}
}



\centerline{
\epsfxsize=8cm\epsfbox{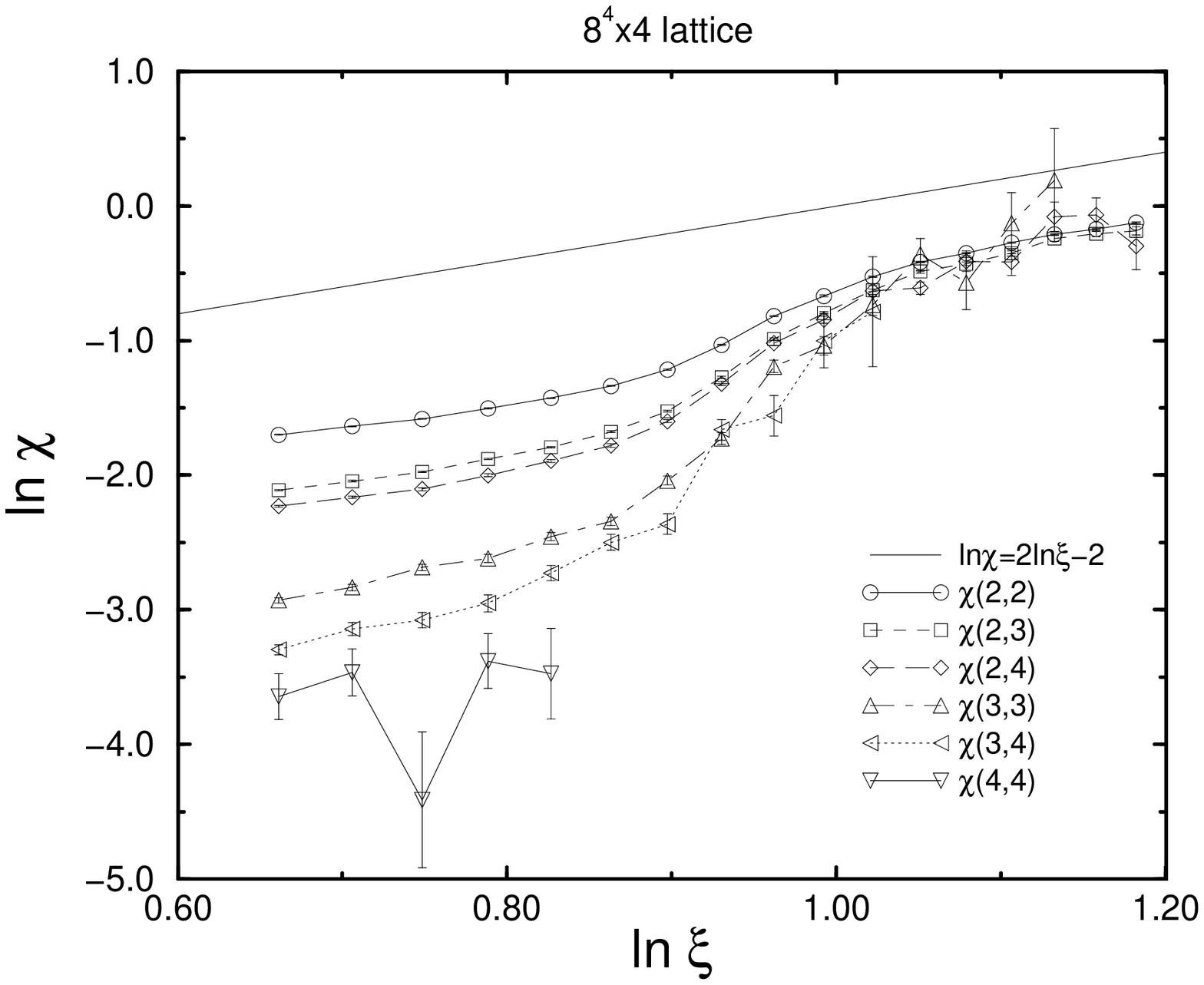}
\epsfxsize=8cm\epsfbox{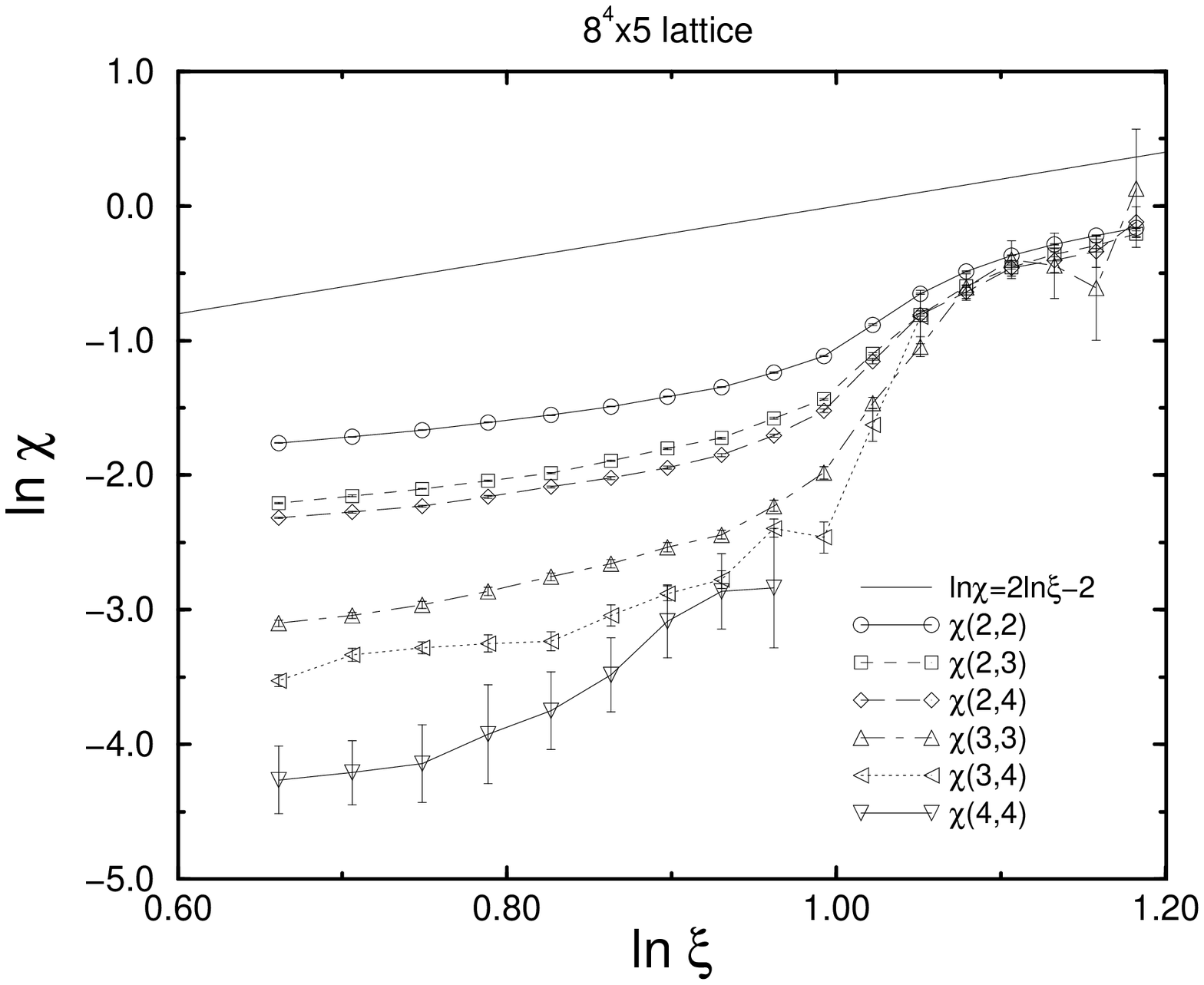}
}



\centerline{
\epsfxsize=8cm\epsfbox{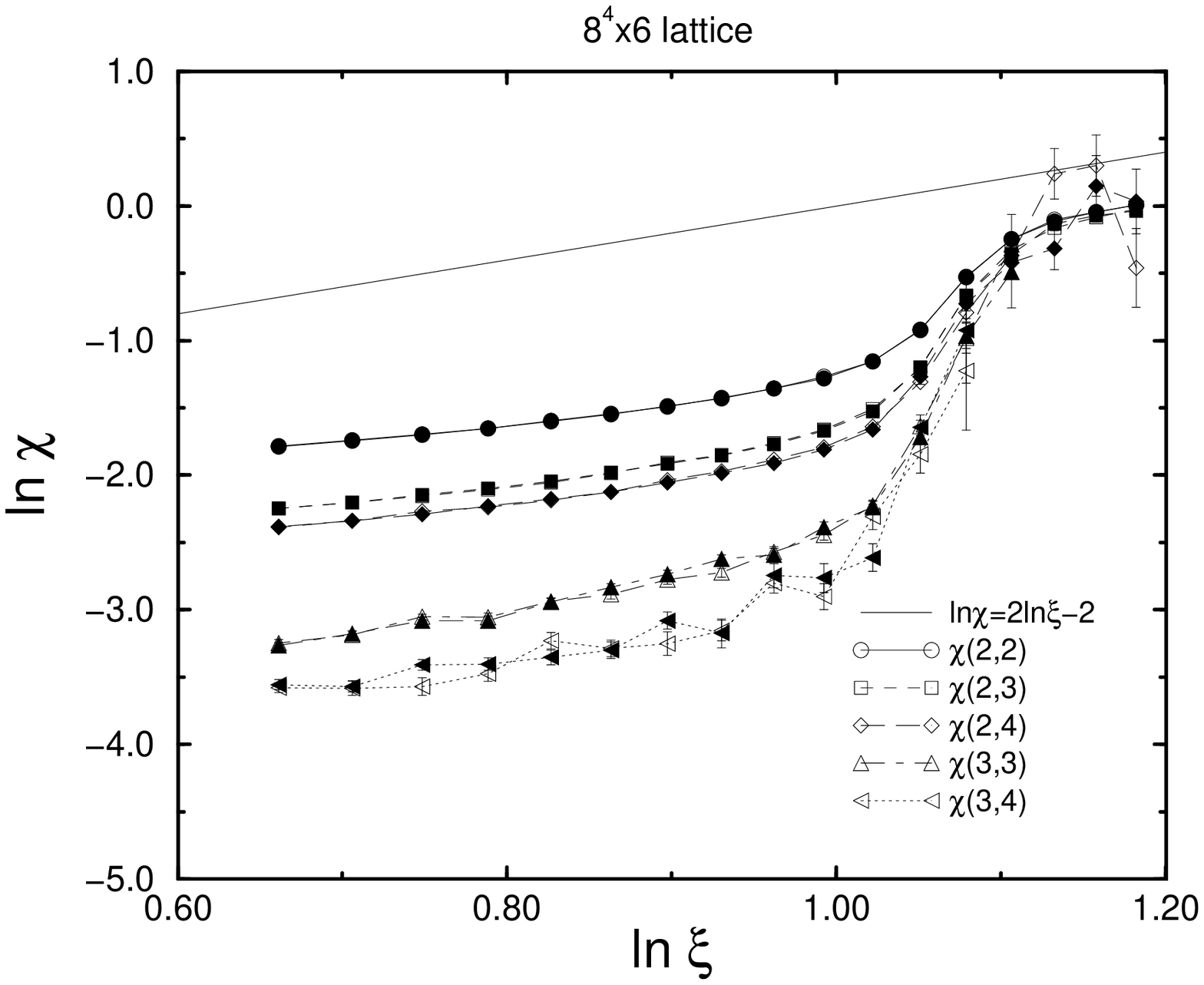}
\epsfxsize=8cm\epsfbox{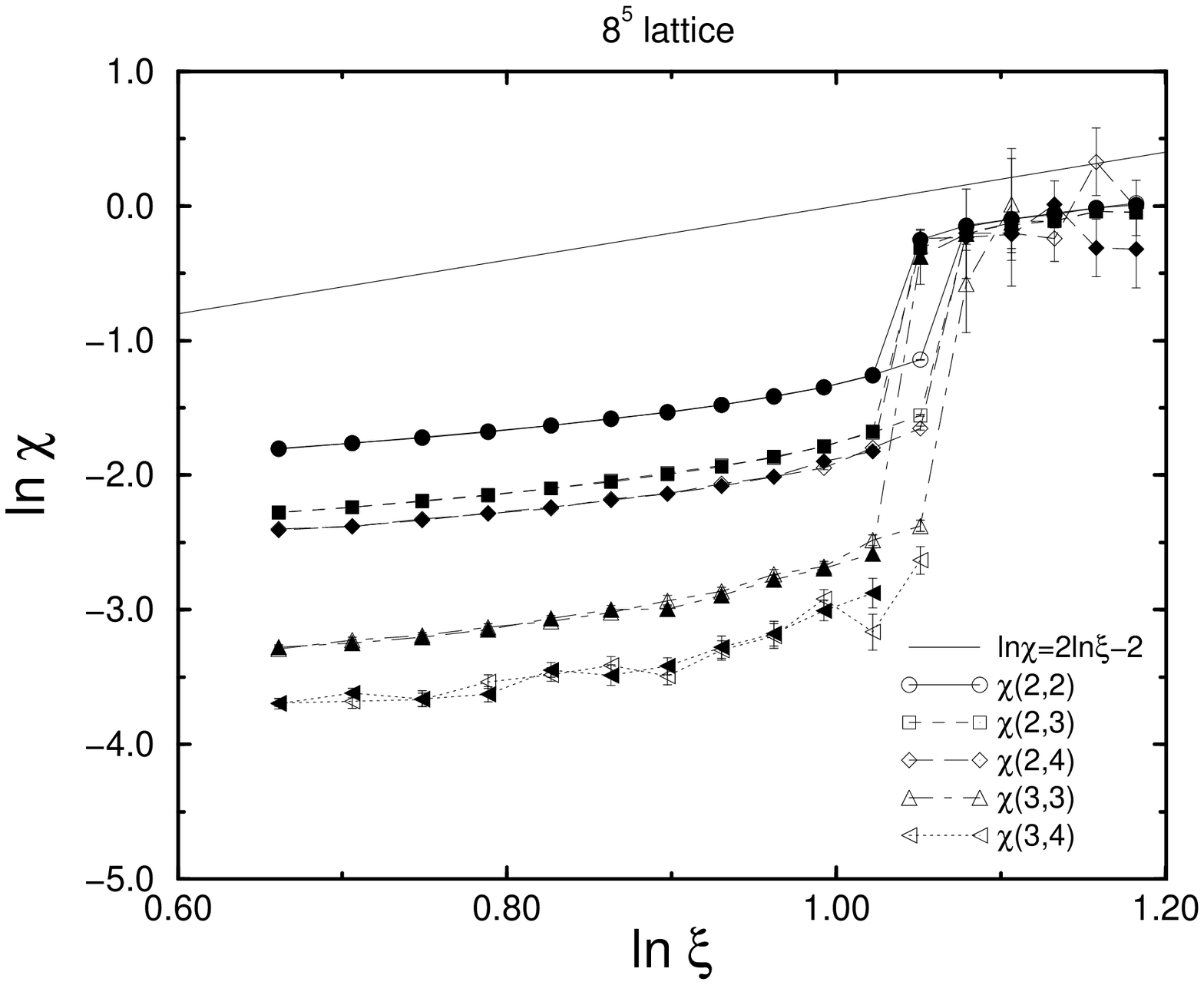}
}


\caption{
Scaling behavior of the Creutz ratio for $N_5=2,3,4,5,6$ and $8$ 
measured on the line 
expected from the one-loop $\beta$-function (\ref{fit2}).
The open symbols are the results of the ordered start and 
the filled symbols are those of the disordered start.
}

\label{fig:creuzscl1}


\end{figure}

\end{document}